\documentclass[twocolumn,english,showpacs,superscriptaddress,prc,aps]{revtex4-1}
\usepackage[T1]{fontenc}
\usepackage[latin9]{inputenc}
\usepackage{babel}
\usepackage{booktabs}
\usepackage{amsmath}
\usepackage{amsthm}
\usepackage{amssymb}
\usepackage{graphicx}
\usepackage{wasysym}
\usepackage[unicode=true,pdfusetitle,
 bookmarks=true,bookmarksnumbered=false,bookmarksopen=false,
 breaklinks=false,pdfborder={0 0 1},backref=false,colorlinks=false]
 {hyperref}
\usepackage{breakurl}
\usepackage{color}

\makeatletter

\providecommand{\tabularnewline}{\\}

\makeatother

\begin{document}

\title{Production of single-charm hadrons by quark combination mechanism in $p$-Pb collisions at $\sqrt{s_{NN}}=5.02$ TeV}

\author{Hai-hong Li}
\affiliation{School of Physics and Engineering, Qufu Normal University, Shandong 273165, China}
\affiliation{Department of Physics, Jining University, Shandong 273155, China} 

\author{Feng-lan Shao}
\email{shaofl@mail.sdu.edu.cn}
\affiliation{School of Physics and Engineering, Qufu Normal University, Shandong 273165, China}

\author{Jun Song }
\email{songjun2011@jnxy.edu.cn}
\affiliation{Department of Physics, Jining University, Shandong 273155, China}

\author{Rui-qin Wang}
\affiliation{School of Physics and Engineering, Qufu Normal University, Shandong 273165, China}

\begin{abstract}
If QGP-like medium is created in $p$-Pb collisions at extremely high collision energies, charm quarks that move in the medium can hadronize by capturing the co-moving light quark(s) or anti-quark(s) to form the charm hadrons. Using light quark $p_{T}$ spectra extracted from the experimental data of light-flavor hadrons and a charm quark $p_{T}$ spectrum that is consistent with perturbative QCD calculations, the central-rapidity data of $p_{T}$ spectra and the spectrum ratios for $D$ mesons in the low $p_{T}$ range ($p_{T}\lesssim7$ GeV/$c$) in minimum-bias $p$-Pb collisions at $\sqrt{s_{NN}}=5.02$ TeV are well described by quark combination mechanism in equal-velocity combination approximation. The $\Lambda_{c}^{+}/D^{0}$ ratio in quark combination mechanism exhibits the typical increase-peak-decrease behavior as the function of $p_{T}$, and the shape of the ratio for $p_{T}\gtrsim3$ GeV/$c$ is in agreement with the data of ALICE collaboration in central rapidity region $-0.96<y<0.04$ and the preliminary data of LHCb collaboration in forward rapidity region $1.5<y<4.0$. The global production of single-charm baryons is quantified using the data and the possible enhancement (relative to light flavor baryons) is discussed. The $p_{T}$ spectra of $\Xi_{c}^{0}$, $\Omega_{c}^{0}$ in minimum-bias events and those of single-charm hadrons in high-multiplicity event classes are predicted, which serves as the further test of the possible change of the hadronization characteristic for low $p_{T}$ charm quarks in the small system created in $p$-Pb collisions at LHC energies. 
\end{abstract}

\maketitle

\section{Introduction}

Quark\textendash gluon plasma (QGP) in which quarks and gluons are deconfined is a new state of the matter of QCD \cite{Shuryak:1980tp}, which is significantly different from the normal nuclear matter. Ultra-relativistic heavy-ion collisions are served as the main experimental approach to study the creation and property of this new state of matter. However, recent ALICE, CMS, and ATLAS experiments find with great surprise that the production of hadrons in high-multiplicity $p$-Pb and $pp$ collisions at LHC exhibits a series of remarkable similarities with that in heavy-ion collisions where QGP is created. These striking observations include long range angular correlations \cite{Khachatryan:2010gv,CMS:2012qk,Aad:2012gla, Aad:2015gqa} and collectivity \cite{Khachatryan:2015waa,Khachatryan:2016txc,Ortiz:2014iva}, enhanced strangeness \cite{Adam:2015vsf,ALICE:2017jyt} and enhanced baryon to meson ratio at low transverse momentum ($p_T$) \cite{Abelev:2013haa,Adam:2016dau}, etc. In heavy ion collisions these phenomena are usually attributed to the creation of QGP. Theoretical explanations on these striking observations in small collision systems usually focus on the creation of mini-QGP or phase transition \cite{Liu:2011dk,Werner:2010ss,Bzdak:2013zma,Bozek:2013uha,Prasad:2009bx,Avsar:2010rf}, multiple parton interaction \cite{Sjostrand:1987su}, color re-connection and string overlap at hadronization \cite{Bautista:2015kwa,Bierlich:2014xba,Ortiz:2013yxa,Christiansen:2015yqa}, etc. In the latest work \cite{Song:2017gcz}, we found that the experimental data of $p_{T}$ spectra for $\phi$, $\Omega^{-}$, $\Xi^{*}\left(1530\right)$, $\text{K}^{*}\left(892\right)$ and other identified hadrons in the low $p_{T}$ range ($p_{T}\lesssim8$ GeV/$c$) in $p$-Pb collisions at $\sqrt{s_{NN}}=5.02$ TeV exhibit an interesting constituent quark number scaling. Such scaling behavior indicates the change of hadronization characteristic in low $p_{T}$ range from the traditional string/cluster fragmentation to the quark (re-)combination mechanism in light (up, down, strange) sector, and is a possible signature for the formation of small dense parton medium in collisions. 

In this paper, we take the $p_{T}$ spectra of single-charm hadrons containing only a charm or anti-charm quark as another probe for the property of the soft parton system created in $p$-Pb collisions at $\sqrt{s_{NN}}=5.02$ TeV, and we study the quark (re-)combination hadronization for low $p_{T}$ charm quarks. This is mainly motivated by the fact that, even though the production of charm quarks is the perturbative QCD process by the initial hard collisions of partons in the incoming nucleons, the hadronization of charm quarks is dependent on the surrounding parton environment. At hadronization, if the charm quark is surrounded by the medium with relatively abundant soft partons, the charm quark can pick up a light antiquark or two light quarks co-moving with it to form a heavy flavor hadron, where the momentum characteristic is the combination $p_{H}=p_{c}+p_{\bar{q},qq}$. Otherwise, in the case of the lack of the co-moving neighbor partons, it will color neutralize by connecting with the faraway parton(s) and fragment into the charm hadron of momentum $p_{H}=xp_{c}$ with $x<1$. This different characteristic of hadronization will be reflected by the $p_{T}$ spectra of charm hadrons in the low $p_{T}$ range. We study such possible change of hadronization characteristic by focusing on the $p_{T}$ spectra of mesons $D^{\pm,0}$, $D_{s}^{+}$ and $D^{*}$, baryons $\Lambda_{c}^{+}$, $\Xi_{c}^{0}$ and $\Omega_{c}^{0}$, and in particular the ratios among them. 

The paper is organized as follows: Sec. II will introduce a working model in quark (re-)combination mechanism for charm quark hadronization.  Sec. III presents our results and relevant discussions. Summary is given at last in Sec. IV. 

\section{charm quark hadronization in QCM}

The (re-)combination of heavy quarks with surrounding light quarks and antiquarks has been suggested in early 1980s \cite{Das:1977cp,Hwa:1979pn,Chiu:1978nc}, and has successfully explained the flavor asymmetry of $D$ mesons at forward rapidities in hadronic collisions through the recombination of charm quarks with valence and/or sea quarks from projectile \cite{Hwa:1994uha,Cuautle:1997ti,Braaten:2002yt, Rapp:2003wn}.  The (re-)combination mechanism is also phenomenologically successful in heavy-ion collisions \cite{Greco:2003vf,vanHees:2005wb,Fries:2008hs,Oh:2009zj,Cao:2013ita,Prino:2016cni,Song:2015ykw,Ghosh:2014oia,Das:2016llg}, where the QGP provides a natural source of thermal light quarks and antiquarks to color neutralize heavy quarks at hadronization. As the aforementioned discussions, if the small dense quark matter is created in $p$-Pb collisions at LHC energies, the low-$p_{T}$ charm quarks will prefer to pick up the co-moving light quarks or antiquarks to form the charm hadrons. In this section, we present a working model for the production of single-charm hadrons in the low $p_{T}$ range in quark (re-)combination mechanism (QCM) in momentum space. 

\subsection{formulism in momentum space}

For a charm meson $M_{c\bar{l}}$ composed of a $c$ and a light antiquark $\bar{l}$, and a charm baryon $B_{cll'}$ composed of a $c$ and two light quarks $l\,l'$, their momentum distributions in QCM, as formulated in e.g. \cite{Wang:2012cw} in general, can be obtained by 
\begin{align}
f_{M_{c\bar{l}}}(p) & =\int dp_{1}dp_{2}f_{c\bar{l}}(p_{1},p_{2})\,{\cal R}_{M_{c\bar{l}}}(p_{1},p_{2};p),\label{eq:fmc}\\
f_{B_{cll'}}(p) & =\int dp_{1}dp_{2}dp_{3}f_{cll'}(p_{1},p_{2},p_{3})\,{\cal R}_{B_{cll'}}(p_{1},p_{2},p_{3};p).\label{eq:fbc}
\end{align}
Here, $f_{c\bar{l}}(p_{1},p_{2})$ is the joint momentum distribution for $c$ and $\bar{l}$. ${\cal R}_{M_{c\bar{l}}}(p_{1},p_{2};p)$ is the combination function that is the probability density for a given $c\bar{l}$ with momenta $p_{1}$, $p_{2}$ combining into a meson $M_{c\bar{l}}$ with momentum $p$. It is similar for baryons.

Considering the perturbative nature of charm quark production, we assume the momentum distribution of charm quarks is independent of those of light quarks. If we also take independent distributions for light quarks of different flavors, we have 
\begin{align}
f_{c\bar{l}}(p_{1},p_{2}) & =N_{c\bar{l}}f_{c}^{\left(n\right)}(p_{1})f_{\bar{l}}^{\left(n\right)}(p_{2}),\label{eq:fcl}\\
f_{cll'}(p_{1},p_{2},p_{3}) & =N_{cll'}f_{c}^{\left(n\right)}(p_{1})f_{l}^{\left(n\right)}(p_{2})f_{l'}^{\left(n\right)}(p_{3}).\label{eq:fcll}
\end{align}
Here we have defined the normalized momentum distribution $f_{c}^{\left(n\right)}\left(p\right)$ with $\int dpf_{c}^{\left(n\right)}\left(p\right)=1$ and $N_{c}$ is charm quark number. It is similar for $f_{l}^{\left(n\right)}\left(p\right)$ and $N_{l}$. The number of quark-antiquark pair reads as $N_{c\bar{l}}=N_{c}N_{\bar{l}}$ and the three-quark combination $N_{cll'}=N_{c}N_{ll'}$ with $N_{ll'}=N_{l}\left(N_{l'}-\delta_{l,l'}\right)$.  $\delta_{l,l'}$ is Kronecker delta function. 

The combination function contains the key information of hadronization.  In sudden hadronization approximation, it is determined by the overlap between the wave function of quarks and that of the hadron or by the Wigner function of the hadron \cite{Fries:2008hs,Greco:2003vf,Chen:2006vc}.  However considering the non-perturbative nature of hadronization, beyond such approximation will be more realistic but in such case we do not know the precise form of the combination function from the solid QCD phenomenology. Therefore, here we only take the most basic characteristic of the combination \textemdash{} the combination mostly happens for quarks and antiquarks that are neighboring in momentum space. We suppose the combination takes place mainly for the quark and/or antiquark which has a given fraction of momentum of the hadron, and we write the combination function
\begin{align}
{\cal R}_{M_{c\bar{l}}}(p_{1},p_{2};p) & =\kappa_{M_{c\bar{l}}}\prod_{i=1}^{2}\delta(p_{i}-x_{i}p),\label{eq:rmc}\\
{\cal R}_{B_{cll'}}(p_{1},p_{2},p_{3};p) & =\kappa_{B_{cll'}}\prod_{i=1}^{3}\delta(p_{i}-x_{i}p),\label{eq:rbc}
\end{align}
where $\kappa_{M_{c\bar{l}}}$ and $\kappa_{B_{cll'}}$ are constants independent of $p$. We adopt the approximation of equal velocity in combination, or called co-moving approximation for heavy quark hadronization. Because the velocity is $v=p/E=p/\gamma m$, equal velocity implies $p_{i}=\gamma vm_{i}\propto m_{i}$ which leads to 
\begin{equation}
x_{i}=m_{i}/\sum_{j}m_{j},
\end{equation}
where quark masses are taken to be the constituent masses in the quark model. Specifically, we take $m_{u}=m_{d}=0.33$ GeV, $m_{s}=0.5$ GeV, and $m_{c}=1.5$ GeV so that the mass and momentum of the hadron can be properly generated by the combination of these constituent quarks and antiquarks. We emphasize that such equal velocity approximation is shown to be quite effective in light sector in our previous work \cite{Song:2017gcz} where the data of $p_{T}$ spectra for identified hadrons such as $p$, $\Lambda$, $\Xi$, $\Omega$, $\phi$, $\text{K}^{*}$, $\Xi^{*}$, and $\Sigma^{*}$ in $p$-Pb collisions at $\sqrt{s_{NN}}=5.02$ TeV can be well explained by a up/down quark spectrum $f_{u}\left(p_{T}\right)$ and a $s$ quark spectrum $f_{s}\left(p_{T}\right)$ at hadronization. For charm hadrons, although the charm quark carries the major part of the momentum of the hadron, light constituent quarks also influence explicitly the momentum distribution of the charm hadron, which can be clearly seen from spectrum ratios such as $D_{s}/D$, $\Lambda_{c}^{+}/D$ and $\Omega_{c}^{0}/D$, etc. 

Substituting Eqs. (\ref{eq:rmc}-\ref{eq:rbc}) and (\ref{eq:fcl}-\ref{eq:fcll}) into Eqs. (\ref{eq:fmc}-\ref{eq:fbc}), we have 
\begin{align}
f_{M_{c\bar{l}}}(p) & =N_{c\bar{l}}\,\kappa_{M_{c\bar{l}}}f_{c}^{\left(n\right)}(x_{1}p)f_{\bar{l}}^{\left(n\right)}(x_{2}p),\label{eq:fmz}\\
f_{B_{cll'}}(p) & =N_{cll'}\kappa_{B_{cll'}}f_{c}^{\left(n\right)}(x_{1}p)f_{l}^{\left(n\right)}(x_{2}p)f_{l'}^{\left(n\right)}(x_{3}p).\label{eq:fbz}
\end{align}
 By defining the normalized meson distribution
\begin{equation}
f_{M_{c\bar{l}}}^{\left(n\right)}\left(p\right)=A_{M_{c\bar{l}}}f_{c}^{\left(n\right)}\left(x_{1}p\right)f_{\bar{l}}^{\left(n\right)}\left(x_{2}p\right),\label{eq:fnmi}
\end{equation}
where $A_{M_{c\bar{l}}}^{-1}=\int dp\,f_{c}^{\left(n\right)}\left(x_{1}p\right)f_{\bar{l}}^{\left(n\right)}\left(x_{2}p\right)$,
and the normalized baryon distribution
\begin{equation}
f_{B_{cll'}}^{\left(n\right)}\left(p\right)=A_{B_{cll'}}\,f_{c}^{\left(n\right)}\left(x_{1}p\right)f_{l}^{\left(n\right)}\left(x_{2}p\right)f_{l'}^{\left(n\right)}\left(x_{3}p\right),\label{eq:fnbi}
\end{equation}
where $A_{B_{cll'}}^{-1}=\int{\rm d}p\prod_{i=1}^{3}f_{q_{i}}^{\left(n\right)}\left(x_{i}p\right)$, we finally obtain the following expressions for charm hadrons 
\begin{align}
f_{M_{c\bar{l}}}\left(p\right) & =N_{M_{c\bar{l}}}\,f_{M_{c\bar{l}}}^{\left(n\right)}\left(p\right),\label{eq:fmfinal}\\
f_{B_{cll'}}\left(p\right) & =N_{B_{cll'}}\,f_{B_{cll'}}^{\left(n\right)}\left(p_{}\right),\label{eq:fbfinal}
\end{align}
with yields 
\begin{align}
N_{M_{c\bar{l}}} & =N_{c\bar{l}}\frac{\kappa_{M_{c\bar{l}}}}{A_{M_{c\bar{l}}}}=N_{c\bar{l}}P_{c\bar{l}\rightarrow M_{c\bar{l}}},  \label{eq:nmi}\\
N_{B_{cll'}} & =N_{cll'}\frac{\kappa_{B_{cll'}}}{A_{B_{cll'}}}=N_{cll'}P_{cll'\rightarrow B_{cll'}}.\label{eq:nbi}
\end{align}
We see that $P_{c\bar{l}\rightarrow M_{c\bar{l}}}\equiv\kappa_{M_{c\bar{l}}}/A_{M_{c\bar{l}}}$ has the proper physical meaning, i.e., the momentum-integrated combination probability for $c\bar{l}\rightarrow M_{c\bar{l}}$. Similarly $P_{cll'\rightarrow B_{cll'}}\equiv\kappa_{B_{cll'}}/A_{B_{cll'}}$ denotes the momentum-integrated combination probability for $cll'\rightarrow B_{cll'}$.

\textcolor{black}{The combination probabilities $P_{c\bar{l}\rightarrow M_{c\bar{l}}}$ and $P_{cll'\rightarrow B_{cll'}}$, correspondingly $\kappa_{B_{cll'}}$ and $\kappa_{M_{c\bar{l}}}$, are parameterized.}  We use $N_{M_{c}}$ to denote the total number of all charm mesons containing one charm constituent. $N_{c\bar{q}}=N_{c}\left(N_{\bar{u}}+N_{\bar{d}}+N_{\bar{s}}\right)$ is the possible number of all charm-light pairs. $N_{M_{c}}/N_{c\bar{q}}$ is then used to estimate the average probability of a $c\bar{q}$ forming a charm meson. For a specific combination $c\bar{l}$, it can have different $J^{P}$ states. In this paper we consider only the pseudo-scalar mesons $J^{P}=0^{-}$($D^{+}$, $D^{0}$ and $D_{s}^{+}$) and vector mesons $J^{P}=1^{-}$ ($D^{*+}$, $D^{*0}$ and $D_{s}^{*+}$) in the ground state. We introduce a parameter $R_{V/P}$ to denote the relative ratio of vector meson to pseudo-scalar meson of the same flavor composition. \textcolor{black}{Then the combination probability of single-charm mesons can be written as 
\begin{equation}
    P_{c\bar{l} \to M_{c\bar{l}}}=C_{M_{c\bar{l}}} \frac{N_{M_{c}}}{N_{c\bar{q}}},\label{eq:pmi}
\end{equation}
with }
\begin{equation}
C_{M_{c\bar{l}}}=\begin{cases}
\frac{1}{1+R_{V/P}} & \text{for}\,J^{P}=0^{-}\,\text{mesons}\\
\frac{R_{V/P}}{1+R_{V/P}} & \text{for}\,J^{P}=1^{-}\,\text{mesons}.
\end{cases}
\end{equation}
By counting polarization states a naive estimation of $R_{V/P}$ is 3.0. However, the mass of the hadron will influence the formation probability in the sense that the lower mass denotes the lower energy level for the bound state formation and means preferable formation. \textcolor{black}{ Here, we roughly estimate such a mass effect through the effective statistical weight and take $R_{V/P}=1.5$, see Appendix \ref{appendixA} for the details and relevant discussions. }

Baryon formation probability $P_{cll'\rightarrow B_{cll'}}$ is obtained similarly. We use $N_{B_{c}}$ to denote the total number of all charm baryons containing one charm quark. $N_{cqq}=N_{c}N_{qq}=N_{c}N_{q}\left(N_{q}-1\right)$ with $N_{q}=N_{u}+N_{d}+N_{s}$ denotes the possible number of all $cqq$ combinations. $N_{B_{c}}/N_{cqq}$ estimates the average probability of the $cqq$ forming a charm baryon. For a specific $cll'$ combination with number $N_{cll'}$, the averaged number of the formed baryons is $N_{iter,ll'}N_{cll'}\frac{N_{B_{c}}}{N_{cqq}}$. Here, $N_{iter,ll'}$ is the iteration number of $ll'$ pair and is taken to be 1 for $l=l'$ and 2 for $l\neq l'$. The appearance of this factor is due to the possible double counting in $N_{qq}$, e.g. $N_{u}N_{d}$ appears twice in $N_{qq}$. We consider the production of single-charm baryons in triplet ($\Lambda_{c}^{+},\Xi_{c}^{+},\Xi_{c}^{0}$) with $J^{P}=\left(1/2\right)^{+}$, in sextet $\left(\Sigma_{c}^{0},\,\Sigma_{c}^{+},\,\Sigma_{c}^{++},\,\Xi_{c}^{'0},\,\Xi_{c}^{'+},\,\Omega_{c}^{0}\right)$ with $J^{P}=\left(1/2\right)^{+}$, and in sextet $\left(\Sigma_{c}^{*0},\,\Sigma_{c}^{*+},\,\Sigma_{c}^{*++},\,\Xi_{c}^{*0},\,\Xi_{c}^{*+},\,\Omega_{c}^{*0}\right)$ with $J^{P}=\left(3/2\right)^{+}$, respectively, in the ground state.  We introduce a parameter $R_{S1/T}$ to denote the relative ratio of $J^{P}=\left(1/2\right)^{+}$ sextet baryons to $J^{P}=\left(1/2\right)^{+}$ triplet baryons of the same flavor composition, and a parameter $R_{S3/S1}$ to denote that of $J^{P}=\left(3/2\right)^{+}$ sextet baryons to $J^{P}=\left(1/2\right)^{+}$ sextet baryons of the same flavor composition.  \textcolor{black}{We also take the effective statistical weight as a guideline and take $R_{S1/T}=0.5$ and $R_{S3/S1}=1.4$, see Appendix \ref{appendixA} for the details and discussions.}

\textcolor{black}{Finally, formation probability $P_{cll'\rightarrow B_{cll'}}$ is written as 
\begin{equation}
P_{cll'\to B_{cll'}}=C_{B_{cll'}}N_{iter,ll'} \frac{N_{B_{c}}}{N_{cqq}},\label{eq:pbi}
\end{equation}
 where $C_{B_{cll'}}$ is the spin-related production weight} according to two parameters $R_{S1/T}$ and $R_{S3/S1}$. For $ll'=uu,dd,ss$, 
\begin{equation}
C_{B_{cll'}}=\begin{cases}
\frac{1}{1+R_{S3/S1}} & \text{for}\,\Sigma_{c}^{++},\Sigma_{c}^{0},\,\Omega_{c}^{0}\\
\frac{R_{S3/S1}}{1+R_{S3/S1}} & \text{for}\,\Sigma_{c}^{*++},\Sigma_{c}^{*0},\,\Omega_{c}^{*0}.
\end{cases}
\end{equation}
For $ll'=ud,us,ds$, 
\begin{equation}
C_{B_{cll'}}=\begin{cases}
\frac{1}{1+R_{S1/T}\left(1+R_{S3/S1}\right)} & \text{for}\,\Lambda_{c}^{+},\,\Xi_{c}^{0},\,\Xi_{c}^{+}\\
\frac{R_{S1/T}}{1+R_{S1/T}\left(1+R_{S3/S1}\right)} & \text{for}\,\Sigma_{c}^{+},\,\Xi_{c}^{'0},\,\Xi_{c}^{'+}\\
\frac{R_{S1/T}R_{S3/S1}}{1+R_{S1/T}\left(1+R_{S3/S1}\right)} & \text{for}\,\Sigma_{c}^{*+},\,\Xi_{c}^{*0},\,\Xi_{c}^{*+}.
\end{cases}
\end{equation}
 We note that after taking the strong and electromagnetic decays into account, yields and momentum spectra of final-state $\Lambda_{c}^{+}$, $\Xi_{c}^{0}$ and $\Omega_{c}^{0}$ which are usually measured in LHC experiments are not sensitive to parameters $R_{S1/T}$ and $R_{S3/S1}$. 

The single-charm mesons and baryons consume most of charm quarks and antiquarks produced in collisions. A rough estimation gives the relative ratios of multi-charm hadrons to single-charm hadrons are only about $N_{M_{c\bar{c}}}/N_{M_{c}}\sim N_{c}/N_{q}\lesssim0.01$ and $N_{Bcc}/N_{Bc}\sim N_{c}/N_{q}\lesssim0.01$.  Therefore, we have the following approximation to single-charm hadrons
\begin{equation}
N_{M_{c}}+N_{B_{c}}\approx N_{c}.\label{eq:nc}
\end{equation}
Here we treat the ratio $R_{B/M}^{\left(c\right)}\equiv N_{B_{c}}/N_{M_{c}}$ as a parameter of the model, which globally characterizes the relative production of single-charm baryons to single-charm mesons. \textcolor{black}{Then we have $N_{M_c} = {N_c}/{(1+R_{B/M}^{(c)})}$ and $N_{B_c}=R_{B/M}^{(c)} N_{M_c}$, and substituting them into Eqs.~(\ref{eq:pmi}) and (\ref{eq:pbi})  we can calculate the yields and momentum distributions of single-charm hadrons through Eqs.~(\ref{eq:fmfinal}) and (\ref{eq:fbfinal}) with a few parameters. }

Some discussions on the present model in contrast with other popular (re-)combination/coalescence models applied in relativistic heavy-ion collisions \cite{Fries:2003vb,Chen:2006vc,Greco:2003vf} are necessary. In essence, our model is a statistical hadronization method based on the constituent quark degrees of freedom, in which unclear non-perturbative dynamics such as the selection of different spin states and the formation competition between baryon and meson in quark combination are treated as model parameters. In addition, it is still unclear at present for the geometrical or spatial structure of the soft parton system in $p$-Pb collisions at LHC, and therefore we do not consider the spatial distributions of quarks at hadronization in the present working model. These points are main difference from those (re-)combination/coalescence models in terms of Wigner function method applied in relativistic heavy-ion collisions \cite{Fries:2003vb,Chen:2006vc,Greco:2003vf}.

\textcolor{black}{Finally, we clarify parameters and/or inputs of the model in study of the single-charm hadrons. Firstly, quark momentum distributions at hadronization are inputs of the model and will be fixed and discussed in the next subsection. Secondly, the values of spin selection parameters $R_{V/P}$, $R_{S1/T}$, $R_{S3/S1}$ are taken from the effective statistical weights. The baryon/meson production competition parameter $R^{(c)}_{B/M}$ is a relatively open parameter and will be discussed in Sec.~\ref{sec3}.}
On the other hand, in the study of the possible creation of deconfined quark matter, results of QCM are usually compared with those of (string) fragmentation mechanism. Because these parameters also exist somehow in string fragmentation, the key phenomenological difference between two classes of hadronization mechanism, in our opinion, lies in the kinetic characteristic in momentum space, which will, for example, obviously exhibit in the ratio of baryons to mesons as the function of $p_{T}$. 

\subsection{$p_{T}$ spectra of constituent quarks at hadronization}

In this paper we study the production of single-charm hadrons at specific rapidities and we apply the formulas in the previous section to the one-dimensional $p_{T}$ space. The $p_{T}$ distributions of quarks and antiquarks at hadronization are inputs of the model. The $p_{T}$ distributions of light constituent quarks in the low $p_{T}$ range are unavailable from the first-principle QCD calculations. However, we can extract them from the data of $p_{T}$ spectra of identified hadrons in QCM in the equal-velocity combination approximation. For example, as formulated in Refs. \cite{Song:2017gcz,Gou:2017foe} which is similar to Eq. (\ref{eq:fmz}), the $p_{T}$ spectrum of $\phi$ is related to that of $s$ quarks

\begin{equation}
f_{\phi}\left(2p_{T}\right)=N_{s\bar{s}}\kappa_{\phi}\left[f_{s}^{\left(n\right)}\left(p_{T}\right)\right]^{2},
\end{equation}
where $\kappa_{\phi}$ is a constant independent of momentum. We can extract $f_{s}^{\left(n\right)}\left(p_{T}\right)$ using the data of $\phi$. $f_{s}^{(n)}\left(p_{T}\right)=f_{\bar{s}}^{(n)}\left(p_{T}\right)$ is assumed at LHC energies due to the charge conjugation symmetry.  The iso-spin symmetry between $u$ and $d$ as well as the charge conjugation symmetry between $u$ and $\bar{u}$ are assumed, so $f_{u}^{\left(n\right)}\left(p_{T}\right)=f_{d}^{\left(n\right)}\left(p_{T}\right)=f_{\bar{u}}^{\left(n\right)}\left(p_{T}\right)=f_{\bar{d}}^{\left(n\right)}\left(p_{T}\right)$, which can be extracted from the spectrum of $\text{K}^{*}$ by the relation
\begin{equation}
f_{K^{*0}}\left(\left(1+r\right)p_{T}\right)=N_{d\bar{s}}\,\kappa_{K^{*0}}\,f_{\bar{s}}^{\left(n\right)}\left(rp_{T}\right)f_{d}^{\left(n\right)}\left(p_{T}\right)
\end{equation}
with the extracted $f_{s}^{\left(n\right)}\left(p_{T}\right)$ and $r=m_{s}/m_{u}$ if the data of $\text{K}^{*}$ are available. Otherwise, it can be extracted from the data of proton after subtracting the decay influence. The number of $s$ quarks and that of $u$ or $d$ quarks are fitted from the data of hadronic yields in QCM. 

\begin{figure}[tbh]
    \includegraphics[width=\linewidth]{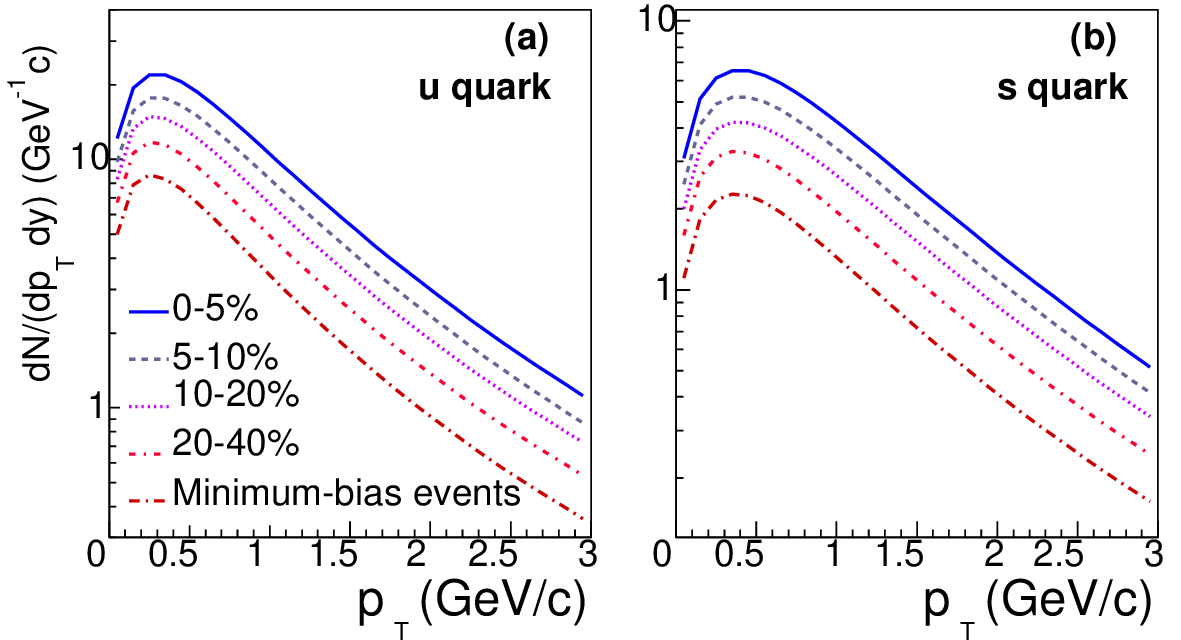}
    \caption{The $p_{T}$ spectra of light quarks at mid-rapidities in $p$-Pb collisions at $\sqrt{s_{NN}}=5.02$ TeV. The $p_{T}$-integrated number densities $dN/dy$ of $u$ and $s$ quarks are (24.6,8.8), (19.7,7.0), (16.0,5.6), (12.2,4.2), and (8.6,2.9) in event classes 0-5\%, 5-10\%, 10-20\%, 20-40\%, and minimum-bias events, respectively. }
    \label{fig1}
\end{figure}

We have obtained these information of light quarks at hadronization in different multiplicity classes in $p$-Pb collisions at $\sqrt{s_{NN}}=5.02$ TeV in previous work \cite{Song:2017gcz}, which is shown in Fig.  \ref{fig1} as the input of the present work. The $p_{T}$ spectra of light quarks in minimum-bias events are also shown.

\begin{figure}[tbh]
    \includegraphics[width=0.8\linewidth]{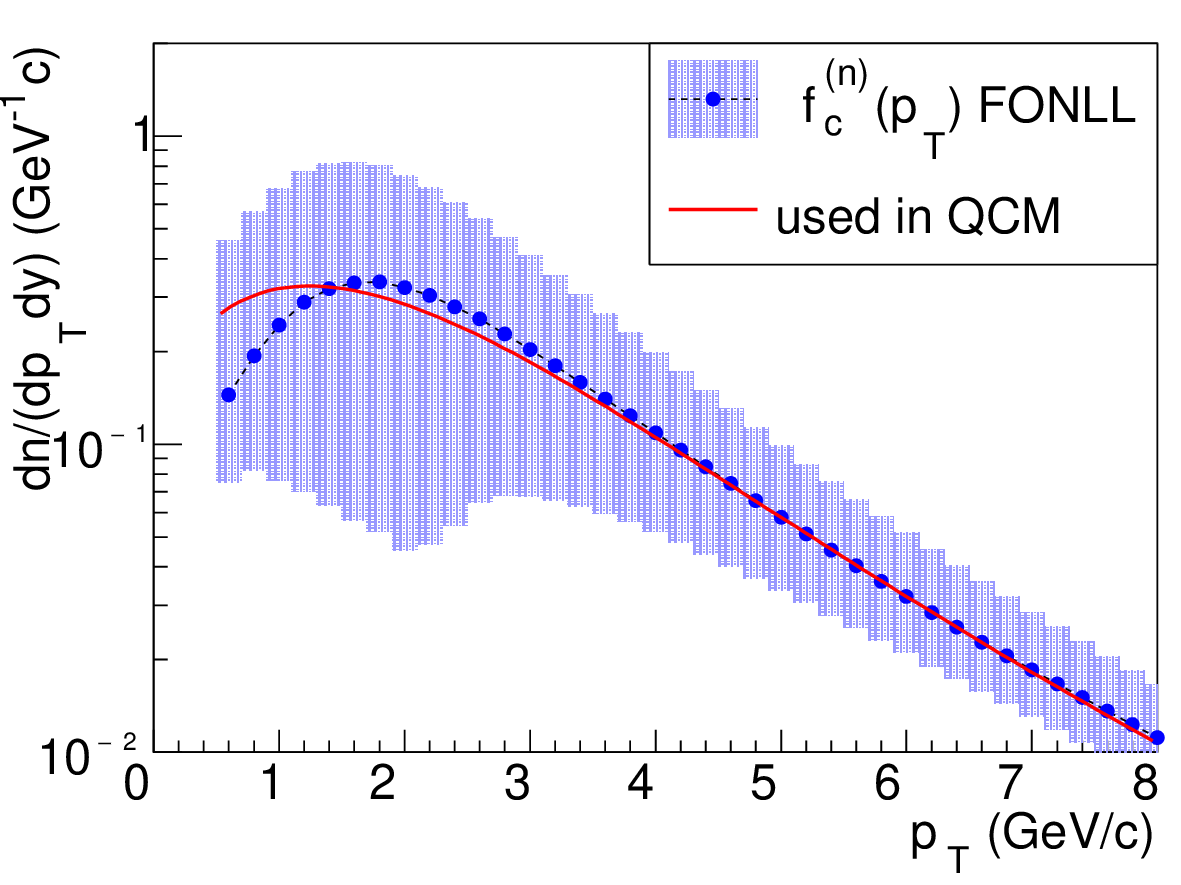}
    \caption{The normalized $p_{T}$ spectrum of charm quarks in rapidity region $-0.96<y<0.04$. The points with the line are center values of FONLL calculation \cite{Cacciari:1998it,Cacciari:2012ny} in $pp$ collisions at $\sqrt{s}=5.02$ TeV and the shadow area shows the theoretical uncertainty. The solid line is the spectrum extracted from the data of $D^{*+}$ mesons \cite{Abelev:2014hha,Adam:2016ich} using QCM. }
    \label{fig2}
\end{figure}

The $p_{T}$ spectrum of charm quarks is calculable in perturbative QCD. Here, we take the calculation of the Fixed-Order Next-to-Leading-Logarithmic (FONLL) \cite{Cacciari:1998it,Cacciari:2012ny} in $pp$ collisions at $\sqrt{s}=5.02$ TeV as the guideline. In Fig. \ref{fig2}, we show the normalized $p_{T}$ distribution of charm quarks in the rapidity interval $-0.96<y<0.04$, which is obtained from the online calculation of FONLL\footnote{FONLL Heavy Quark Production, http://www.lpthe.jussieu.fr/\textasciitilde{}cacciari/fonll/fonllform.html}.  The points with the line are center values of FONLL and the shadow area shows the scale uncertainties, see Refs. \cite{Cacciari:1998it,Cacciari:2012ny} for details. The uncertainty of parton distribution functions (PDFs) is not included. We see that the theoretical uncertainty is large for the spectrum of charm quarks at low $p_{T}$. If we directly use this spectrum, our results of charm hadrons also have large uncertainties and the comparison with the data will be less conclusive. Therefore we only take the calculation of FONLL as an important guideline. The practical $p_{T}$ spectrum of charm quarks used in this paper is extracted by fitting the data of $D^{*+}$ mesons in rapidity region $-0.96<y<0.04$ \cite{Abelev:2014hha,Adam:2016ich} in minimum-bias $p$-Pb collisions at $\sqrt{s_{NN}}=5.02$ TeV in QCM and is shown as the thick solid line in Fig. \ref{fig2}. We see that the extracted spectrum is very close to the center points of FONLL calculation for $p_{T}\gtrsim1.5$ GeV/$c$ and excesses the latter to a certain extent for lower $p_{T}$. 

\section{Comparison with data and discussions}\label{sec3}

In this section, we use the above working model to describe the available data of the $p_{T}$ spectra of $D$ mesons and $\Lambda_{c}^{+}$ baryon in central and forward rapidity regions in minimum-bias $p$-Pb collisions at $\sqrt{s_{NN}}=5.02$ TeV. We study to what extent these data are described by QCM, and discuss what information on the hadronization of low $p_{T}$ charm quarks can be extracted from these data. We give the prediction of other single-charm baryons $\Xi_{c}^{0}$ and $\Omega_{c}^{0}$ as well as the $p_{T}$ spectra of $D$ mesons, $\Lambda_{c}^{+}$, $\Xi_{c}^{0}$ and $\Omega_{c}^{0}$ baryons in different multiplicity classes in $p$-Pb collisions at $\sqrt{s_{NN}}=5.02$ TeV for the further test of the model. 

\subsection{$D$ meson spectra and spectrum ratios}

\begin{figure}[tbh]
    \includegraphics[width=\linewidth]{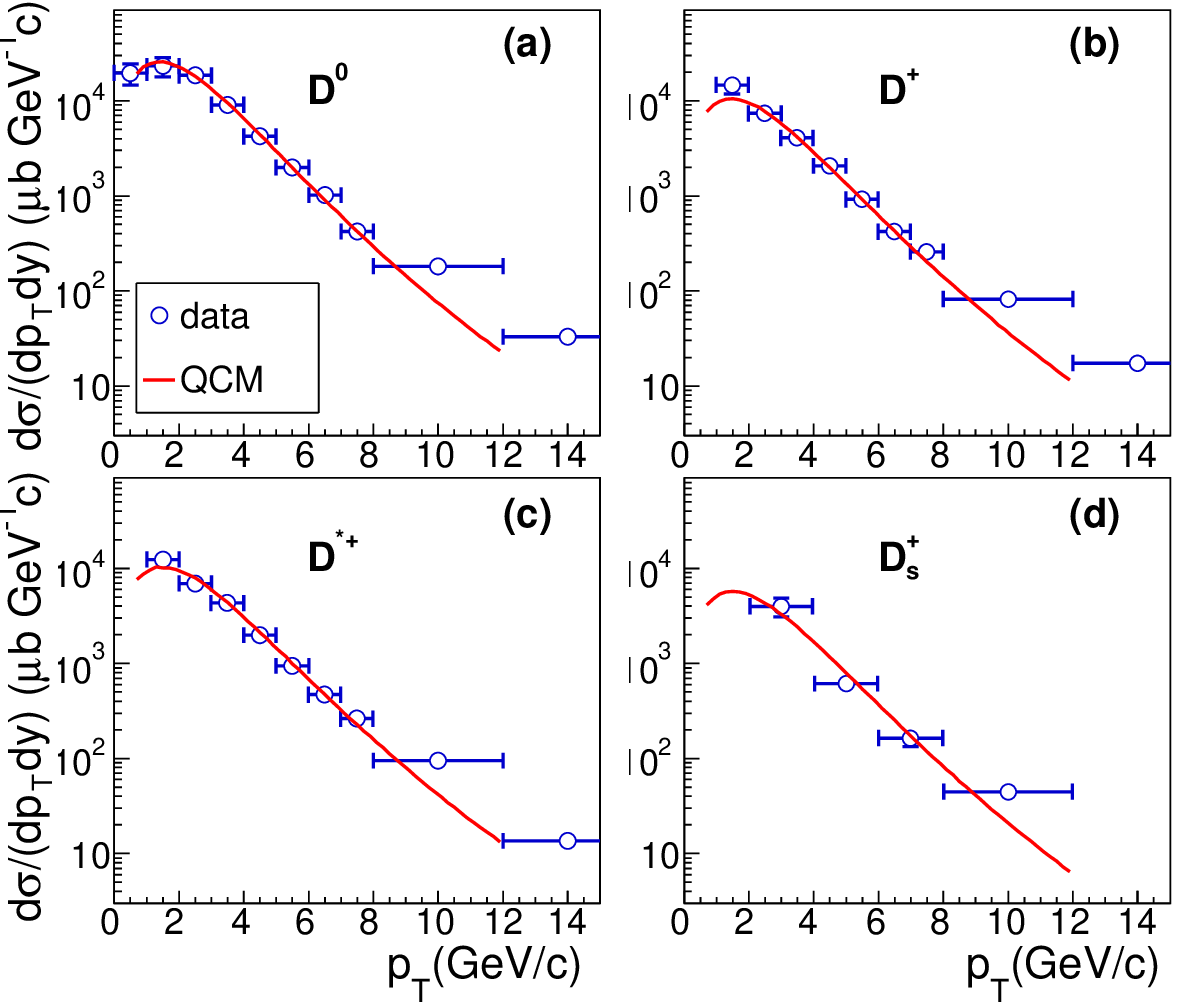}
    \caption{Differential cross sections of $D$ mesons as the function of $p_{T}$ in central rapidity region $-0.96<y<0.04$ in minimum-bias $p$-Pb collisions at $\sqrt{s_{NN}}=5.02$ TeV. Symbols are experimental data \cite{Abelev:2014hha,Adam:2016ich} and lines are results of QCM.}
    \label{fig3}
\end{figure}

In Fig. \ref{fig3}, we show the differential cross sections of $D$ mesons as the function of $p_{T}$ in central rapidity region $-0.96<y<0.04$ in minimum-bias $p$-Pb collision at $\sqrt{s_{NN}}=5.02$ TeV. Symbols are experimental data \cite{Abelev:2014hha,Adam:2016ich} and lines are results of QCM. The data of $D^{*+}$ are used to determine the $p_{T}$ spectrum of charm quarks at hadronization. The normalized charm quark spectrum $f_{c}^{\left(n\right)}\left(p_{T}\right)$ was shown in Fig. \ref{fig2}. The $p_{T}$-integrated cross section of charm quarks $d\sigma_{c}/dy$ can be extracted from the $D^{*+}$ cross section by
\begin{align}
    \frac{d\sigma_{c}}{dy} & =\left(2+\lambda_{s}\right)\left(1+R_{B/M}^{\left(c\right)}\right)\left(1+\frac{1}{R_{V/P}}\right)\frac{d\sigma_{D^{*+}}}{dy}. \label{eq:sigmac1}
\end{align}
Here, $\lambda_{s}\equiv\frac{d\sigma_{s}}{dy}/\frac{d\sigma_{u}}{dy}=\frac{dN_{s}}{dy}/\frac{dN_{u}}{dy}$ is the strangeness suppression factor and is $0.34\pm0.01$ in $p$-Pb collisions \cite{Shao:2017eok}. $R_{V/P}$ is taken to be the thermal-weight value 1.5. 
\textcolor{black}{The fitted ${d\sigma_{D^{*+}}}/{dy}$ is about 32 mb. Parameter $R_{B/M}^{(c)}$ is taken to be 0.425 according to the study of $\Lambda_c^{+}$ in the next subsection. The resulting $d\sigma_{c}/dy$ is 178 mb. We find that it is higher than the experimental extraction $151\pm14(\text{stat})_{-26}^{+13}(\text{syst})$ mb using the data of $D^0$ \cite{Adam:2016ich} and the fraction of charm quarks hadronizing into $D^0$ mesons $f_{c\to D^{0}}=0.542\pm0.024$ \cite{Gladilin:2014tba}. This overestimation is because the fraction $f_{c\to D^{0}}=0.422$ in our model at $R_{B/M}^{(c)}=0.425$ is smaller than the experimentally used fraction. The calculated $d\sigma_{D^{0}}/dy=75.0$ mb in our model is consistent with the experimental data $79.0\pm7.3\left(\textrm{stat}\right)_{-13.4}^{+7.1}\left(\textrm{syst}\right)$ mb \cite{Adam:2016ich}. 
}


In Fig. \ref{fig3}, we see that results of QCM are in good agreement with the data for $p_{T}\lesssim7$ GeV/$c$ but are smaller than the data for high momenta $p_{T}\gtrsim8$ GeV/$c$. This is reasonable. Supposing the existence of the parton medium with ample (dozens of) quarks and antiquarks with soft momenta $p_{T_{l}}\lesssim2$ GeV/$c$, during  moving in the medium the perturbatively-created charm quark with momentum $p_{T,c}\lesssim5$ GeV/$c$ has many potential co-moving light antiquarks and it can pick up one of them to form the single-charm meson. For the hadron formation at high momentum $p_{T}\gtrsim8$ GeV/$c$, if it hadronizes still by combination, a charm quark with $p_{T,c}\gtrsim6$ GeV/$c$ should find out or occur the co-moving light antiquark with $p_{T,l}\gtrsim2$ GeV/$c$ which number drops rapidly. In this case the combination is not the dominated channel and the fragmentation will take over. 

\begin{figure}[tbh]
    \includegraphics[width=\linewidth]{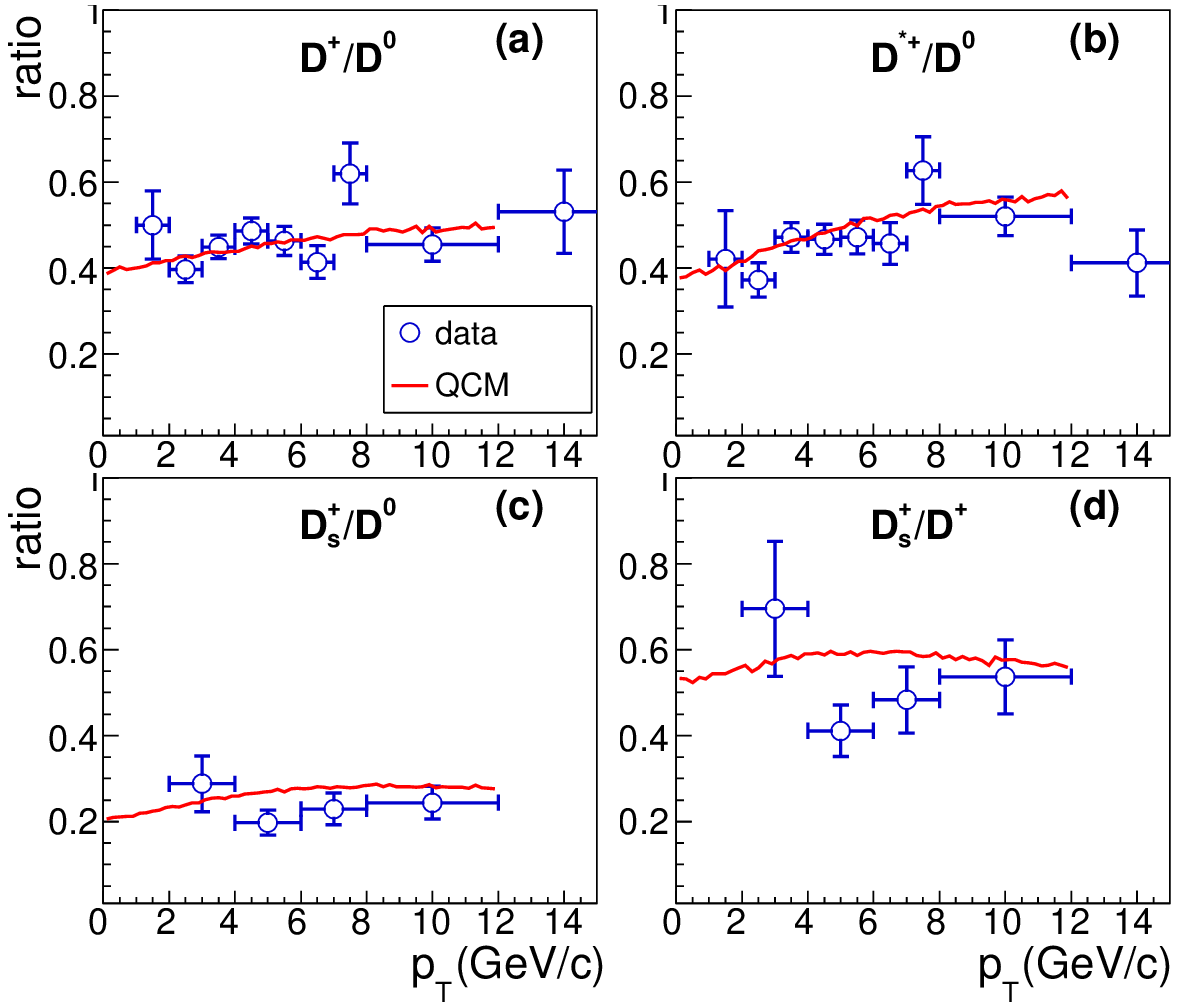}
    \caption{Ratios of differential cross sections for $D$ mesons as the function of $p_{T}$ in central rapidity region $-0.96<y<0.04$ in minimum-bias $p$-Pb collisions at $\sqrt{s_{NN}}=5.02$ TeV. Experimental data are taken from \cite{Abelev:2014hha,Adam:2016ich}.  The sawtooth of QCM results is due to the finite statistics.}
    \label{fig4}
\end{figure}
In Fig. \ref{fig4}, we show results for the ratios of differential cross sections for $D$ mesons as the function of $p_{T}$. Symbols are experimental data \cite{Abelev:2014hha,Adam:2016ich} and lines are results of QCM. The sawtooth behavior in our results is due to the finite statistics. Our results of $D^{+}/D^{0}$, $D^{*+}/D^{0}$ and $D_{s}^{+}/D^{0}$ are in good agreement with the data. The comparison for $D_{s}^{+}/D^{+}$ is not conclusive. 

The $p_{T}$-integrated cross sections $d\sigma/dy$ and the averaged transverse momenta $\langle p_{T}\rangle$ of $D$ mesons are shown in Table \ref{tab1}. Several ratios for the cross sections of charm hadrons are interesting. Using Eq. (\ref{eq:nmi}) and taking the strong and electromagnetic decay contribution into account, we obtain the ratios of cross sections for $D$ mesons
\begin{equation}
\frac{D^{+}}{D^{0}}=\frac{1+0.323R_{V/P}}{1+1.677R_{V/P}}\approx0.42,\label{eq:ratio_Dp_D0}
\end{equation}

\begin{align}
\frac{D^{*+}}{D^{0}} & =\frac{R_{V/P}}{1+1.677R_{V/P}}\approx0.43,\label{eq:ratio_Dstar_D0}
\end{align}
\begin{equation}
\frac{D_{s}^{+}}{D^{0}}=\frac{1+R_{V/P}}{1+1.677R_{V/P}}\lambda_{s}\approx0.24,\label{eq:ratio_Ds_D0}
\end{equation}

\begin{equation}
\frac{D_{s}^{+}}{D^{0}+D^{+}}=\frac{1}{2}\lambda_{s}\approx0.17,
\end{equation}
where we have used $\lambda_{s}=0.34\pm0.01$ in $p$-Pb collisions \cite{Shao:2017eok} and $R_{V/P}=1.5$. We see that results of $D^{+}/D^{0}$, $D^{*+}/D^{0}$ and $D_{s}^{+}/D^{0}$ are consistent with the observation in low $p_{T}$ range in Fig. \ref{fig4}. In addition, ratio $D_{s}^{+}/\left(D^{0}+D^{+}\right)$ is only dependent on the $\lambda_{s}$ in QCM and therefore is a potentially interesting measurement. 

\begin{table}
\caption{The $p_{T}$-integrated cross sections $d\sigma/dy$ and $\langle p_{T}\rangle$
of $D$ mesons in minimum-bias $p$-Pb collisions at $\sqrt{s_{NN}}=5.02$
TeV. }

\begin{tabular*}{7cm}{@{\extracolsep{\fill}}ccccc}
\toprule 
 & $D^{0}$ & $D^{+}$ & $D_{s}^{+}$ & $D^{*+}$\tabularnewline
\midrule
\midrule 
$\frac{d\sigma}{dy}$(mb) & 75.0 & 31.2 & 17.7 & 32.0\tabularnewline
\midrule 
$\langle p_{T}\rangle$(GeV/$c$) & 2.23 & 2.29 & 2.36 & 2.35\tabularnewline
\bottomrule
\end{tabular*}\label{tab1}
\end{table}

\textcolor{black}{We note that the experimental data of $p_{T}$ spectra and spectrum ratios of $D$ mesons in $p$-Pb collisions at $\sqrt{s_{NN}}=5.02$ TeV can be explained by perturbative-QCD calculations with fragmentation functions \cite{Kramer:2017gct,Acharya:2017jgo} within the theoretical uncertainties.  In contrast, our results indicate another kind of the understanding for these data.} We emphasize two points in our method (1) the mid-rapidity $p_{T}$ spectra of light-flavor quarks in the low $p_{T}$ range are extracted from the data of light-flavor hadrons using quark combination mechanism; (2) the used $p_{T}$ spectrum of charm quarks is consistent with perturbative-QCD calculations.  Therefore, our results suggest a possibly universal new picture for the production of light and heavy flavor hadrons in low $p_{T}$ range at the hadronization of the small parton system in $p$-Pb collisions at LHC energies. 

\subsection{$\Lambda_{c}^{+}/D^{0}$ ratio as the function of $p_{T}$}

The production of baryons is sensitive to the hadronization mechanism.  In particular, the ratio of baryon to meson as the function of $p_{T}$ is usually served as an effective probe of the hadron production mechanism.  The enhancement of the baryon/meson ratios at ``intermediate'' $p_{T}$ ($2\apprle p_{T}\apprle5$ GeV/$c$) is a characteristic behavior of QCM.  In experiments, the enhancement of the ratios for light-flavor hadrons such as $p/\pi$, $\Lambda/\text{K}_{s}^{0}$ and $\Omega/\phi$ has been observed many times in relativistic heavy-ion collisions \cite{Abelev:2013vea,Abelev:2013xaa,Abelev:2014uua,Abelev:2006jr,Abelev:2007rw}, and it is also recently observed in $p$-Pb collisions at LHC energies \cite{Abelev:2013haa}. The data of the $\Lambda_{c}^{+}/D^{0}$ ratio in central rapidity region $-0.96<y<0.04$ in $p$-Pb collisions at $\sqrt{s_{NN}}=5.02$ TeV are recently reported by ALICE collaboration, \textcolor{black}{ which  show an enhancement \cite{Acharya:2017kfy}.}  In addition, LHCb collaboration also presents the preliminary data of $\Lambda_{c}^{+}/D^{0}$ ratio at forward and backward rapidities, which show the similar enhancement and further indicate the nontrivial $p_{T}$ dependence of the ratio. In this subsection, we use QCM to understand these recent data for the $\Lambda_{c}^{+}/D^{0}$ ratio. 

\begin{figure}
    \includegraphics[width=\linewidth]{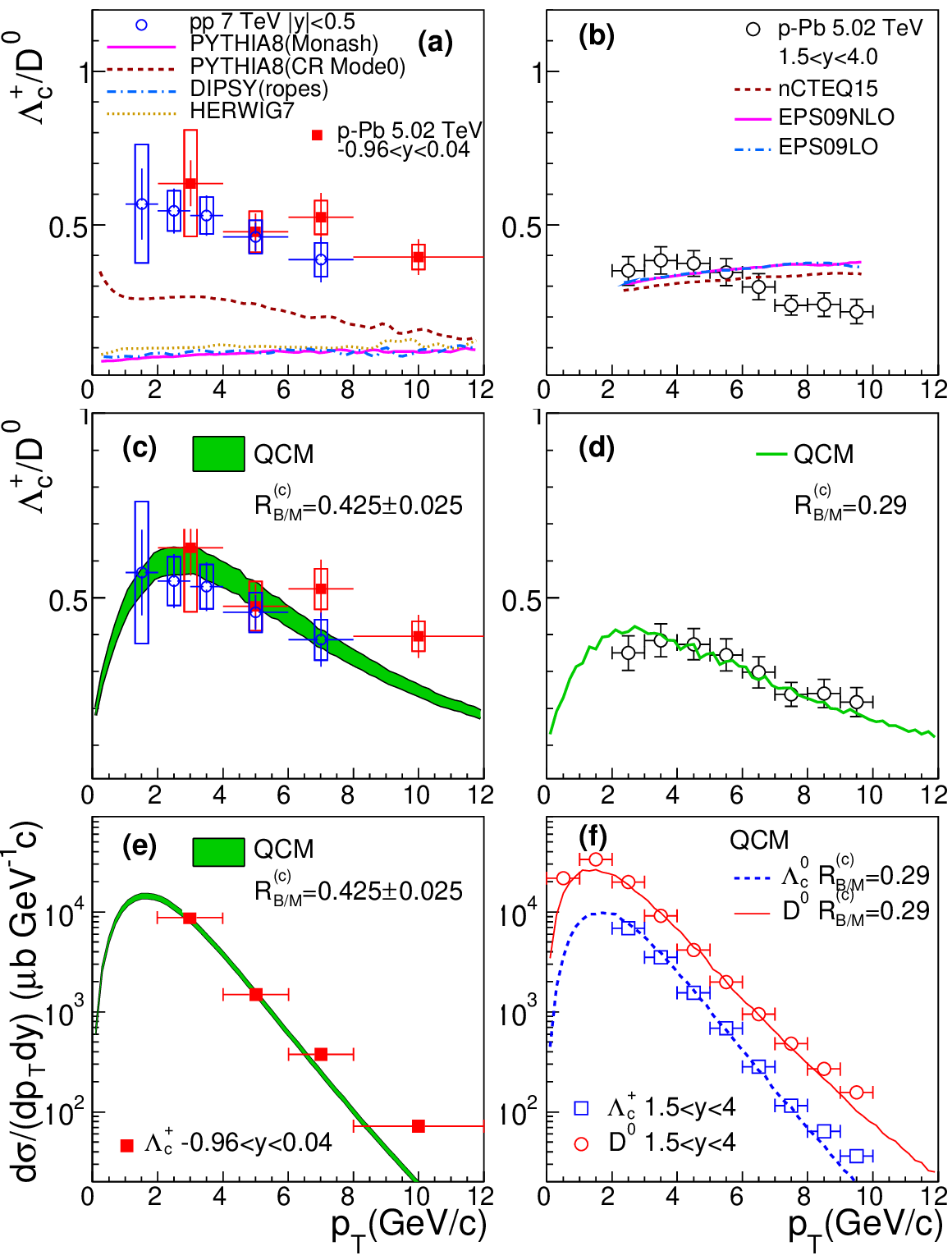}
    \caption{The $\Lambda_{c}^{+}/D^{0}$ ratio as the function of $p_{T}$ in minimum-bias $p$-Pb collisions at $\sqrt{s_{NN}}=5.02$ TeV. Symbols are experimental data of ALICE and LHCb collaborations in central and forward rapidity regions \cite{Acharya:2017kfy,LHCb:2017rvh}. Lines in panels (a) and (b) are results of models or event generators which adopt the fragmentation hadronization and are taken from \cite{Acharya:2017kfy,LHCb:2017rvh}.  Lines (with band) in panels (c-f) are QCM results. }
    \label{fig5}
\end{figure}

Firstly, we briefly discuss the experimental data of the $\Lambda_{c}^{+}/D^{0}$ ratio at LHC in central and forward rapidity regions and some existing theoretical predictions. The data of $\Lambda_{c}^{+}/D^{0}$ ratio in central rapidity region $-0.96<y<0.04$ in $p$-Pb collisions at $\sqrt{s_{NN}}=5.02$ TeV \cite{Acharya:2017kfy}, solid squares, are shown in Fig. \ref{fig5}(a). \textcolor{black}{Here, the data of $pp$ collisions at midrapidity ($|y|<0.5$) at $\sqrt{s}=7$ TeV, open circles, are also shown and are consistent with the data of $p$-Pb collisions, which implies the similarity of charm quark hadronization in $pp$ and $p$-Pb collisions at LHC energies \cite{Song:2018tpv}.} These data show the ratio $\Lambda_{c}^{+}/D^{0}$ at LHC energies reaches about 0.5 in $p_{T}$ range ($2\lesssim p_{T}\lesssim5$ GeV/$c$) and seems to decrease with the $p_{T}$ as $p_{T}\gtrsim2$ GeV/$c$. The predictions from popular event generators PYTHIA8 \cite{Sjostrand:2007gs}, DIPSY \cite{Bierlich:2014xba}, and HERWIG \cite{Bahr:2008pv} are also shown in Fig. \ref{fig5}(a) with different kinds of lines. These event generators adopt the string or cluster fragmentation to describe the hadronization. We see that predictions of DIPSY, HERWIG and PYTHIA8 without color re-connection give the $\Lambda_{c}^{+}/D^{0}$ ratio of about 0.1 which is significantly smaller than the data and give a slightly increasing tendency with $p_{T}$. Considering the color re-connection effects in PYTHIA8 \cite{Christiansen:2015yqa} can increase the ratio up to about 0.3 and give the decreasing tendency with $p_{T}$. In Fig. \ref{fig5}(b), the preliminary data of $\Lambda_{c}^{+}/D^{0}$ ratio in forward rapidity region $1.5<y<4.0$ in $p$-Pb collisions at $\sqrt{s_{NN}}=5.02$ TeV are shown. The data in forward rapidity region show the $\Lambda_{c}^{+}/D^{0}$ ratio in $p_{T}$ range ($2\lesssim p_{T}\lesssim5$ GeV/$c$) is about 0.36 which is smaller than that in central rapidity region. The data show the $\Lambda_{c}^{+}/D^{0}$ ratio decreases with $p_{T}$ as $p_{T}\gtrsim3$ GeV/$c$ but the first data point at 2.5 GeV/$c$ is relatively smaller than the peak and therefore may indicate the possible decrease of the ratio at smaller $p_{T}.$ NLO theoretical predictions \cite{Lansberg:2016deg,Shao:2015vga} with parton distribution functions EPS09LO, EPS09NLO and nCTEQ15 are shown in Fig.\ref{fig5}(b) as different kinds of lines, which have the proper magnitude for the ratio in $p_{T}$ range ($2\lesssim p_{T}\lesssim5$ GeV/$c$) but give the slightly increasing tendency with $p_{T}$. 

After the above discussions, we conclude that two features of the $\Lambda_{c}^{+}/D^{0}$ ratio, i.e., the magnitude of the ratio in $2\lesssim p_{T}\lesssim5$ GeV/$c$ range and the $p_{T}$ dependence of the ratio (i.e., ratio shape), are key test for the theoretical models. In QCM, the baryon/meson ratio usually exhibits an enhancement at ``intermediate'' $p_{T}$ and the $p_{T}$ dependence of the ratio is non-monotonic, i.e., the ratio increases at small $p_{T}$ and peaks at $p_{T}\sim2-3$ GeV/$c$ and then decreases at larger $p_{T}$.  For the $p_{T}$ dependence of the $\Lambda_{c}^{+}/D^{0}$ ratio, we can see this typical behavior from the ratio of directly produced $\Lambda_{c}^{+}$ to $D^{0}$. The extracted $p_{T}$ spectra of light quarks and charm quarks at hadronization in minimum-bias events shown in Figs. \ref{fig1} and \ref{fig2} can be parameterized by the following form \footnote{We adopt this kind of parameterization instead of the usual form based on L\'evy-Tsallis parameterization in order to better separate the term increasing the baryon/meson ratio from the decrease one. }
\begin{equation}
f_{q}^{\left(n\right)}\left(p_{T}\right)=\mathcal{A}_{q}\sqrt{p_{T}}\exp\left[P_{4,q}\left(p_{T}\right)\right],\label{eq:fqPara}
\end{equation}
where $P_{4,q}\left(p_{T}\right)$ is the polynomial of degree four and is taken to be 
\begin{equation}
P_{4,u}\left(p_{T}\right)=-2.098p_{T}-0.1508p_{T}^{2}+0.1807p_{T}^{3}-0.02684p_{T}^{4}
\end{equation}
 for $u$ quarks in $p_{T}$ range {[}0,3{]} GeV/$c$ and 
\begin{equation}
P_{4,c}\left(p_{T}\right)=-0.02778p_{T}-0.1919p_{T}^{2}+0.02197p_{T}^{3}-0.0008231p_{T}^{4}
\end{equation}
 for $c$ quarks in $p_{T}$ range {[}0,10{]} GeV/$c$, respectively. $\mathcal{A}_{q}$ is the normalization. Following Eqs. (\ref{eq:fmz}) and (\ref{eq:fbz}), the ratio of directly produced $\Lambda_{c}^{+}$ to directly produced $D^{0}$ is
\begin{align}
\frac{f_{\Lambda_{c}^{+}}\left(p_{T}\right)}{f_{D^{0}}\left(p_{T}\right)} & =\frac{N_{cud}\kappa_{\Lambda_{c}^{+}}f_{c}^{\left(n\right)}\left(\frac{m_{c}\,p_{T}}{m_{c}+2m_{u}}\right)\left[f_{u}^{\left(n\right)}\left(\frac{m_{u}\,p_{T}}{m_{c}+2m_{u}}\right)\right]^{2}}{N_{c\bar{u}}\kappa_{D^{0}}f_{c}^{\left(n\right)}\left(\frac{m_{c}\,p_{T}}{m_{c}+m_{u}}\right)f_{u}^{\left(n\right)}\left(\frac{m_{u}\,p_{T}}{m_{c}+m_{u}}\right)}\nonumber \\
 & \approx1.15\frac{N_{\Lambda_{c}^{+}}}{N_{D^{0}}}\sqrt{p_{T}}\,e^{-\frac{p_{T}\left(1-\frac{p_{T}}{19.1}\right)\left(\left(\frac{p_{T}-3.02}{8.15}\right)^{2}+1\right)}{4.38}}.\label{eq:rLD}
\end{align}
We see that the $p_{T}$ dependence of the ratio is controlled by two terms. The first term is $\sqrt{p_{T}}$ which increases the ratio with $p_{T}$.  The second term is the exponential term which monotonically decreases with $p_{T}$. The competition between two terms leads to that the ratio increases at small $p_{T}$ and then reaches the peak at $p_{T}\approx3$ GeV/$c$ and decreases for larger $p_{T}$. \textcolor{black}{We emphasize that this non-monotonic $p_T$ dependence of the $\Lambda_{c}^{+}/D^{0}$ ratio essentially comes from the non-monotonic shape of quark distribution $dn/dp_Tdy$, see Figs. \ref{fig1} and \ref{fig2}. The $p_{T}$ spectrum of baryon is the product of $p_{T}$ spectra of three quarks, which is the third power of quark's non-monotonic shape with $p_{T}$. The $p_{T}$ spectrum of meson is the product of that of quark and that of antiquark, which is the square of quark's non-monotonic shape with $p_{T}$. Therefore, the $\Lambda_{c}^{+}/D^{0}$ ratio has the non-monotonic $p_{T}$ dependence and it is a general property of baryon/meson ratio in QCM. } In addition, taking the decay contribution into account only slightly changes the shape of the ratio. 

In heavy-ion collisions, the light quark spectra at low $p_{T}$ exhibit the thermal feature, i.e. $f_{l}^{(n)}\left(p_{T}\right)=dn/dp_{T}\propto p_{T}\exp[-\sqrt{p_{T}^{2}+m^{2}}/T]$.  Therefore, the first term that controls the increase of the baryon/meson ratio is $p_{T}$ instead of $\sqrt{p_{T}}$ and the ratio can reach higher peak value at ``intermediate'' $p_{T}$ in heavy-ion collisions, which had been observed for light-flavor hadrons such as $p/\pi$, $\Lambda/K_{s}^{0}$ and $\Omega/\phi$ \cite{Abelev:2013vea,Abelev:2013xaa,Abelev:2014uua,Abelev:2006jr,Abelev:2007rw}. 

The $\Lambda_{c}^{+}/D^{0}$ spectrum ratio in Eq. (\ref{eq:rLD}) is also influenced by the yield ratio $N_{\Lambda_{c}^{+}}/N_{D^{0}}$ which describes the global production of $\Lambda_{c}^{+}$ relative to $D^{0}$. We illustrate this point by the ratio of the $p_{T}$-integrated cross section of $\Lambda_{c}^{+}$ to that of $D^{0}$. Using Eqs. (\ref{eq:nmi}) and (\ref{eq:nbi}) and taking the strong and electromagnetic decays into account we have
\begin{equation}
\frac{\Lambda_{c}^{+}}{D^{0}}=\frac{4\frac{1}{2+\lambda_{s}}}{\frac{1+1.677R_{V/P}}{1+R_{V/P}}}R_{B/M}^{\left(c\right)}=1.215R_{B/M}^{\left(c\right)},\label{eq:ratio_Lambda_c_D0}
\end{equation}
where $\lambda_{s}=0.34$ and $R_{V/P}=1.5$ are taken. We see that the ratio is directly influenced by the parameter $R_{B/M}^{\left(c\right)}$ which quantifies the production competition of single-charm baryons to mesons at hadronization. Because of the difficulty of non-perturbative QCD, $R_{B/M}^{\left(c\right)}$ can not be determined by first-principle calculations and here we treat it as the parameter of the model.

The bands in Fig. \ref{fig5} (c) and (e) are our results of $\Lambda_{c}^{+}/D^{0}$ ratio and $\Lambda_{c}^{+}$ spectrum as the function of $p_{T}$ using light quark spectra in Fig. \ref{fig1} and the charm quark spectrum in Fig. \ref{fig2} with $R_{B/M}^{\left(c\right)}=0.425\pm0.025$.  We see that the shape of the ratio and that of $\Lambda_{c}^{+}$ spectrum in $3\lesssim p_{T}\lesssim7$ GeV/$c$ range are consistent with the data in central rapidity region $-0.96<y<0.04$. We also use a $u$ quark spectrum $f_{u}\left(p_{T}\right)$ and a charm quark spectrum $f_{c}\left(p_{T}\right)$ to fit the data of $p_{T}$ spectra of $\Lambda_{c}^{+}$ and $D^{0}$ at forward rapidities $1.5<y<4.0$ and results with $R_{B/M}^{\left(c\right)}=0.29$ are shown in Fig.  \ref{fig5} (f). Here we note that the extracted $f^{(n)}_{c}\left(p_{T}\right)$ at the forward rapidity is also very close to the center values of FONLL calculation  \cite{Cacciari:1998it,Cacciari:2012ny}. The result of $\Lambda_{c}^{+}/D^{0}$ ratio in forward rapidity region $1.5<y<4$ is shown in Fig.  \ref{fig5} (d). We see that the data of $p_{T}$ spectra of $\Lambda_{c}^{+}$ and $D^{0}$ and their ratio in $3\lesssim p_{T}\lesssim7$ GeV/$c$ range can be well fitted by QCM. 

In addition, we find that, except the difference in global magnitude due to $R_{B/M}^{\left(c\right)}$, the shape of $\Lambda_{c}^{+}/D^{0}$ ratio with respect to $p_{T}$ in forward rapidity region $1.5<y<4.0$ is very close to that in central rapidity region $-0.96<y<0.04$ in our model. This is because of the following reasons which can be seen from Eq. (\ref{eq:rLD}). Since the charm quark carries most of momenta of the $\Lambda_{c}^{+}$ and $D^{0}$ in QCM, the change of charm quark spectrum at different rapidities is largely canceled in $\Lambda_{c}^{+}/D^{0}$ ratio. In addition, the momenta of light quarks that take part in the formation of $\Lambda_{c}^{+}$ and $D^{0}$ are usually $p_{T}\lesssim2$ GeV/$c$. The change of light quark spectra for $p_{T}\lesssim2$ GeV/$c$ at different rapidities is not significant and is only weakly passed to the ratio because the light (anti-)quarks carry small fraction of momenta of the $\Lambda_{c}^{+}$ and $D^{0}$.  This is different from that in light-flavor hadrons where the baryon/meson ratios have nontrivial dependence on the change of quark spectra because there each quark carries similar fraction of the momenta of the formed hadrons due to the similar constituent quark masses. 

An important result in the above fitting is that we extract $R_{B/M}^{\left(c\right)}=0.425\pm0.025$ in central rapidity region and $R_{B/M}^{\left(c\right)}=0.29$ in forward rapidity region. The parameter $R_{B/M}^{\left(c\right)}$ characterizes the relative production of single-charm baryons to single-charm mesons as the charm quark hadronizes. There are two striking properties for the extracted $R_{B/M}^{\left(c\right)}$. First, we see a large difference between $R_{B/M}^{\left(c\right)}$ in central rapidity region and that in forward rapidity region. This is somewhat puzzling.  For light-flavor hadrons, the ratios of baryons to mesons such as $p/\pi$ and $\Lambda/K_{s}^{0}$ yield ratios at LHC energies are relatively stable at different system sizes and in different rapidity regions \cite{Abelev:2013xaa,Abelev:2013vea,Khachatryan:2011tm,Khachatryan:2016yru,ALICE:2017jyt,Shao:2017eok}.  Models or event generators based on string or cluster fragmentation also usually predict the stable baryon/meson yield ratios with the rapidity at LHC \cite{Bierlich:2014xba,Christiansen:2015yqa,Lansberg:2016deg}.  In addition, the preliminary data of LHCb collaboration \cite{LHCb:2017rvh} show $\Lambda_{c}^{+}/D^{0}$ ratio is stable at different forward and backward rapidities. Second, the extracted $R_{B/M}^{\left(c\right)}$ is larger than that extrapolated from light-flavor hadrons. If we suppose that the baryon/meson production competition in the formation of single-charm hadrons is the same as that in the formation of light-flavor hadrons, we have the relative ratio $R_{B/M}^{\left(c\right)}\approx3R_{B/M}^{\left(l\right)}=0.26$, \textcolor{black}{see Appendix \ref{appendixB} for the derivation of this relation}.  Here, $R_{B/M}^{\left(l\right)}\equiv N_{B(lll)}/N_{M(l\bar{l})}$ is the ratio of light-flavor baryons to mesons and is about 0.086 by fitting the data of light-flavor hadrons in our previous work \cite{Shao:2017eok}.  This extrapolated value of $R_{B/M}^{\left(c\right)}\approx0.26$ is close to (10\% lower than) the extracted $R_{B/M}^{\left(c\right)}$ in forward rapidity region but is significantly (60\%) lower than that in central rapidity region. On the other hand, the $\Lambda_{c}^{+}/D^{0}$ ratio as $R_{B/M}^{\left(c\right)}\approx0.26$ is about 0.32, see Eq. (\ref{eq:ratio_Lambda_c_D0}), which is close to the thermal estimation $\Lambda_{c}^{+}/D^{0}\simeq0.25-0.3$ by statistical hadronization model \cite{Kuznetsova:2006bh,Andronic:2010dt}. In addition, we note that the parameter $R_{V/P}=1.5$ in meson production which can well describe the data of $D$ meson ratios in Fig. \ref{fig4} is also from the thermal weight. 

The following discussions are helpful to understand the large value for the extracted $R_{B/M}^{\left(c\right)}$. First is the naive estimation from the stochastic color combination. When the charm quark moves in the medium, supposing the colors of neighboring light quarks and/or antiquarks are stochastic, then the probability of the charm quark with specific color (e.g. \emph{R}) occurring the light anti-quark with right anti-color ($\bar{R}$) to form the color singlet is 1/3, and the probability of this charm quark occurring two light quarks with right colors (corresponding to $\bar{R}$) to form the color singlet is 1/9. Then we have $R_{B/M}^{\left(c\right)}=\frac{1}{9}/\frac{1}{3}=1/3$ by the stochastic color combination, which is close to the extraction in the forward rapidity region. Second is the possible effect of the correlated color combination. As the charm quark hadronizes, the surrounding medium (those light quarks and antiquarks) is also in the vicinity of the hadronization. The colors of two light quarks neighboring in phase space will tend to be $\bar{3}$ states that have attractive force.  Then in the single-charm baryon formation the probability of the charm quark occurring two light quarks with right colors can be greater than 1/9 (the stochastic one) and $R_{B/M}^{\left(c\right)}>1/3$. Such possible enhancement has been studied by considering the existence of diquark in QGP near hadronization in Refs. \cite{Sateesh:1991jt,Lee:2007wr,Oh:2009zj}, which can raise the yield ratio $\Lambda_{c}^{+}/D^{0}>1.3$ (corresponding to parameter $R_{B/M}^{\left(c\right)}\gtrsim1$ in our model). In addition, the wave functions of charm hadrons will also influence the combination probability, for example, in several quark coalescence models through hadronic Wigner function \cite{Fries:2003vb,Greco:2003vf,Chen:2006vc}, which suggests the significantly enhanced ratio $\Lambda_{c}^{+}/D^{0}>0.8$ in relativistic heavy-ion collisions \cite{Oh:2009zj, Plumari:2017ntm} (corresponding to parameter $R_{B/M}^{\left(c\right)}>0.7$ in our model). 

We finally give a short summary of this subsection. The $\Lambda_{c}^{+}/D^{0}$ ratio in our model exhibits the typical behavior of baryon/meson ratios as the function of $p_{T}$ in quark combination mechanism. The shape of the calculated $\Lambda_{c}^{+}/D^{0}$ ratio is consistent with the experimental data in central and forward rapidity regions in $p$-Pb collisions at $\sqrt{s_{NN}}=5.02$ TeV \cite{Acharya:2017kfy,LHCb:2017rvh}.  We suggest the measurement of the more precise data points, especially those for $p_{T}\lesssim3$ GeV/$c$, in order to better test the quark combination characteristic for charm quark hadronization and quantify the enhancement of $\Lambda_{c}^{+}$ baryon in low $p_{T}$ range in $p$-Pb collisions at LHC energies. 

\subsection{predictions for other hadrons and multiplicity classes}

\begin{figure}[t]
    \includegraphics[width=\linewidth]{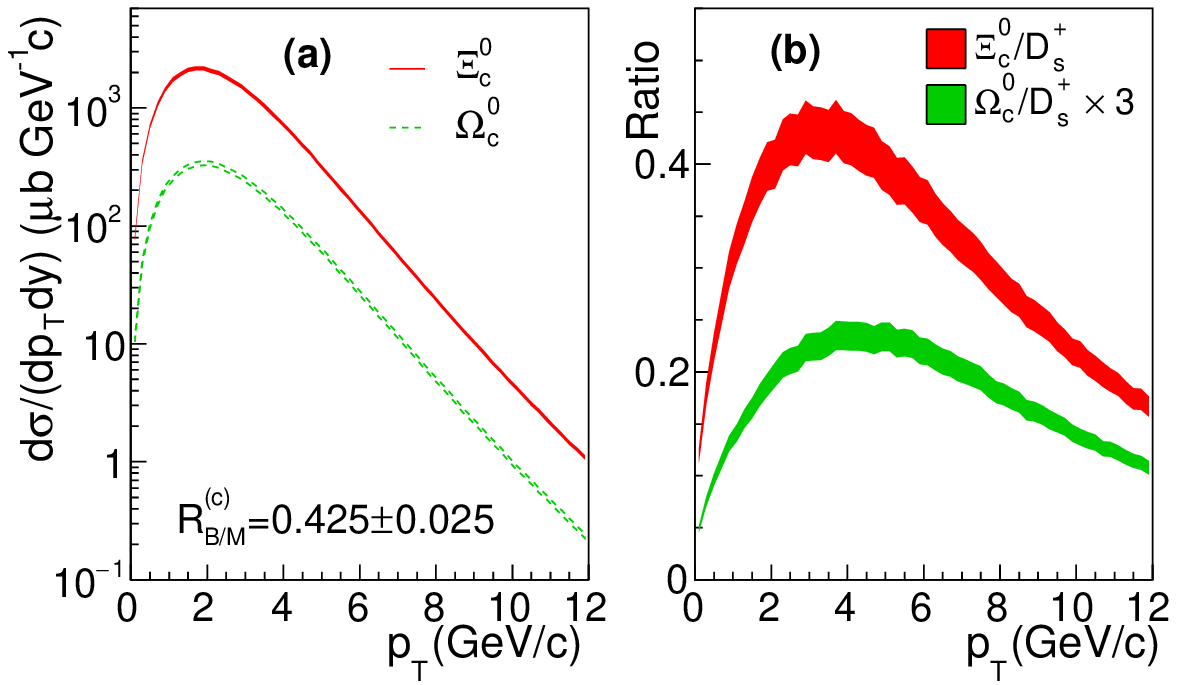}
    \caption{The prediction of $p_{T}$ spectra of $\Xi_{c}^{0}$ and $\Omega_{c}^{0}$ and their ratios to $D_{s}^{+}$ in central rapidity region $-0.96<y<0.04$ in minimum-bias $p$-Pb collisions at $\sqrt{s_{NN}}=5.02$ TeV. The uncertainties of $p_{T}$ spectra of $\Xi_{c}^{0}$ and $\Omega_{c}^{0}$ due to that of $R_{B/M}^{\left(c\right)}=0.425\pm0.025$ in panel (a) are too small to be visible. }
    \label{fig6}
\end{figure}

The production of other single-charm baryons such as $\Xi_{c}^{0}$ and $\Omega_{c}^{0}$ will also exhibit the similar enhancement in QCM. In Fig. \ref{fig6}, we predict the $p_{T}$ spectra of $\Xi_{c}^{0}$ and $\Omega_{c}^{0}$ and their ratios to $D_{s}^{+}$ in central rapidity region $-0.96<y<0.04$ in minimum-bias $p$-Pb collisions at $\sqrt{s_{NN}}=5.02$ TeV. The parameter $R_{B/M}^{\left(c\right)}=0.425\pm0.025$ is taken according to the discussion of $\Lambda_{c}^{+}$ production in previous subsection. We see that both $\Xi_{c}^{0}/D_{s}^{+}$ and $\Omega_{c}^{0}/D_{s}^{+}$ ratios exhibit the typical increase-peak-decrease behavior as the function of $p_{T}$. In addition, we note that the magnitude of $\Xi_{c}^{0}/D_{s}^{+}$ at peak position $p_{T}\approx3$ GeV/$c$ is about 0.45 which is slightly smaller than that of $\Lambda_{c}^{+}/D^{0}$ at similar peak position. $\Omega_{c}^{0}/D_{s}^{+}$ ratio has the much lower magnitude than $\Xi_{c}^{0}/D_{s}^{+}$ ratio because of the strangeness suppression. The peak position of $\Omega_{c}^{0}/D_{s}^{+}$ ratio is about $p_{T}\approx4$ GeV/$c$ which is about 1 GeV/$c$ larger than those of $\Lambda_{c}^{+}/D^{0}$ and $\Xi_{c}^{0}/D_{s}^{+}$ ratios.  This is because the $p_{T}$ spectrum of strange quarks is flatter than that of up/down quarks \cite{Song:2017gcz}. 

The above results are of minimum-bias events in $p$-Pb collisions which have relatively small charged-particle multiplicity $dN_{ch}/d\eta=17.6$ at midrapidity \cite{ALICE:2012xs}. In fact, as indicated by the observation of collectivity and strangeness enhancement \cite{Khachatryan:2015waa,ALICE:2017jyt,Adam:2015vsf}, QGP-like medium is most probably created in high-multiplicity events of $p$-Pb collisions at LHC, and QCM should be preferably applied in these events. Results of QCM for light-flavor hadrons have shown this point \cite{Song:2017gcz}. Therefore, we make predictions for $p_{T}$ spectra of single-charm hadrons in high-multiplicity events for the future test. The selected multiplicity classes are I(0-5\%), II(5-10\%), III(10-20\%) and IV(20-40\%), which are four highest multiplicity classes in measurements of ALICE collaboration. The cross sections of charm quarks $d\sigma_{c}/dy$ in the four multiplicity classes are fixed in QCM by fitting the experimental data of the multiplicity dependence of $D$ meson cross sections \cite{Adam:2016mkz}. With $R_{B/M}^{\left(c\right)}=0.425\pm0.025$, they are taken to be ($594\pm35$, $448\pm26$, $360\pm21$, $254\pm15$) mb in four classes, respectively.  We neglect the possible effects of energy loss for charm quarks traveling in the medium before hadronization and use the extracted spectrum of charm quarks in the minimum-bias events. 

\begin{figure*}[tbh]
    \includegraphics[width=0.98\linewidth]{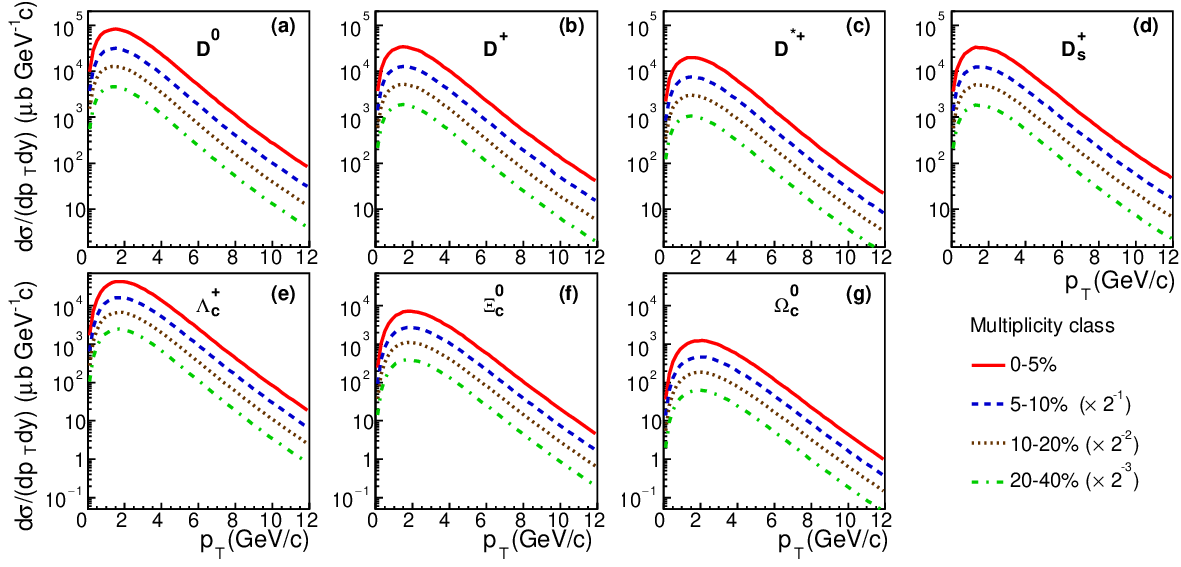}
    \caption{The prediction of $p_{T}$ spectra of single-charm hadrons in central rapidity region $-0.96<y<0.04$ in different multiplicity classes in $p$-Pb collisions at $\sqrt{s_{NN}}=5.02$ TeV. }
    \label{fig7}
\end{figure*}

In Fig. \ref{fig7}, we show $p_{T}$ spectra of single-charm hadrons in central rapidity region in different multiplicity classes in $p$-Pb collisions at $\sqrt{s_{NN}}=5.02$ TeV. In Table \ref{tab2}, we show the $p_{T}$-integrated cross sections and $\langle p_{T}\rangle$ of single-charm hadrons for $R_{B/M}^{\left(c\right)}=0.425$. We see that the $\langle p_{T}\rangle$ of charm hadrons slightly increase with multiplicity and the increase of baryons is slightly larger than that of mesons. This is due to the increased $\langle p_{T}\rangle$ of light quarks in high-multiplicity events. 

\begin{table}
\caption{The $p_{T}$-integrated cross sections $d\sigma/dy$ and $\langle p_{T}\rangle$
of single-charm hadrons in central rapidity region $-0.96<y<0.04$
in different multiplicity classes in $p$-Pb collisions at $\sqrt{s_{NN}}=5.02$
TeV. The four multiplicity classes are I(0-5\%), II(5-10\%), III(10-20\%)
and IV(20-40\%), respectively.}

\begin{tabular*}{8cm}{@{\extracolsep{\fill}}ccccccccc}
\toprule 
 & \multicolumn{4}{c}{$\frac{d\sigma}{dy}$(mb)} & \multicolumn{4}{c}{$\langle p_{T}\rangle$(GeV/$c$)}\tabularnewline
\midrule 
class & I & II & III & IV & I & II & III & IV\tabularnewline
\midrule
\midrule 
$D^{0}$ & 246 & 185 & 149 & 106 & 2.32 & 2.31 & 2.28 & 2.25\tabularnewline
\midrule 
$D^{+}$ & 103 & 77.8 & 62.7 & 44.4 & 2.38 & 2.37 & 2.34 & 2.31\tabularnewline
\midrule 
$D_{s}^{+}$ & 61.9 & 46.7 & 37.1 & 25.5 & 2.41 & 2.40 & 2.39 & 2.35\tabularnewline
\midrule 
$D^{*+}$ & 104 & 78.8 & 63.5 & 45.0 & 2.44 & 2.43 & 2.40 & 2.36\tabularnewline
\midrule 
    $\Lambda_{c}^{+}$ & 127 & 96.4 & 77.8 & 55.4 & 2.45 & 2.44 & 2.39 & 2.34\tabularnewline
\midrule 
$\Xi_{c}^{0}$ & 22.7 & 17.1 & 13.6 & 9.40 & 2.62 & 2.61 & 2.57 & 2.51\tabularnewline
\midrule 
$\Omega_{c}^{0}$ & 4.02 & 3.03 & 2.38 & 1.59 & 2.73 & 2.72 & 2.70 & 2.63\tabularnewline
\bottomrule
\end{tabular*}

\label{tab2}
\end{table}

\begin{figure*}[tbh]
    \includegraphics[width=0.98\linewidth]{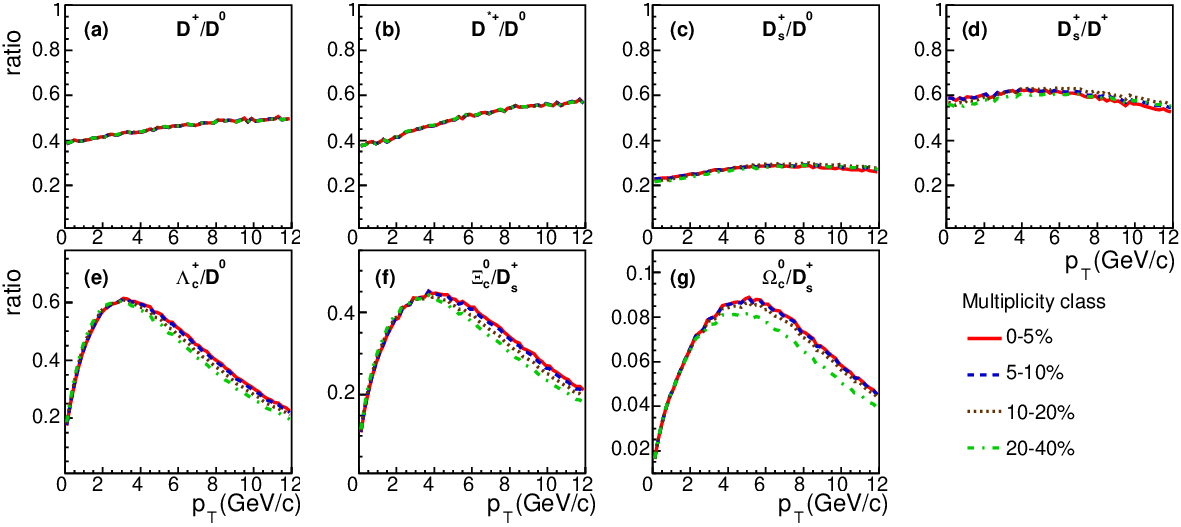}
    \caption{The prediction of ratios among different single-charm hadrons as the function of $p_{T}$ in different multiplicity classes in $p$-Pb collisions at $\sqrt{s_{NN}}=5.02$ TeV. The lines are results with $R_{B/M}^{\left(c\right)}=0.425$. }
    \label{fig8}
\end{figure*}

In Fig. \ref{fig8}, we show the ratios among different single-charm hadrons as the function of $p_{T}$ in different multiplicity classes.  We see that ratios $D^{+}/D^{0}$ and $D^{*+}/D^{0}$ are hardly changed in different multiplicity classes. We first note that their $p_{T}$-integrated ratios Eqs. (\ref{eq:ratio_Dp_D0}) and (\ref{eq:ratio_Dstar_D0}) are independent of the multiplicity. Even though the $p_{T}$ spectrum of up/down quarks is weakly changed in different multiplicity classes, as shown in Fig. \ref{fig1}, this change will influence the inclusive spectra of $D^{+}$, $D^{*+}$ and $D^{0}$ but is significantly canceled in their ratios. As indicated by Eq. (\ref{eq:ratio_Ds_D0}), spectrum ratios $D_{s}^{+}/D^{0}$ and $D_{s}^{+}/D^{+}$ are explicitly dependent on the strangeness suppression factor $\lambda_{s}$. Because the change of $\lambda_{s}$ is only about 0.02 in the studied four multiplicity classes \cite{Song:2017gcz}, the spectrum ratios $D_{s}^{+}/D^{0}$ and $D_{s}^{+}/D^{+}$ also change little. The multiplicity dependence of ratios $\Lambda_{c}^{+}/D^{0}$, $\Xi_{c}^{0}/D^{+}_{s}$ and $\Omega_{c}^{0}/D^{+}_{s}$ is more obvious than that of meson ratios. Such increasing dependence is because the formation of single-charm baryon needs one  more light quark than that of single-charm meson and therefore the baryon/meson ratios suffer more influence of the change of light quark spectra. 

\section{Summary}

We have studied the production of single-charm hadrons in $p$-Pb collisions at $\sqrt{s_{NN}}=5.02$ TeV in quark combination mechanism.  Considering the possible creation of the dense parton medium in $p$-Pb collisions at such extremely-high collision energies, the charm quark hadronizes by capturing the co-moving light quark or antiquark in dense medium to form the single-charm hadron. We introduced a working model in the framework of quark combination mechanism to formulate the yields and momentum distributions of single-charm hadrons formed by the combination of charm quarks and light-flavor (anti-)quarks in equal-velocity combination approximation. The data of $p_{T}$ spectra of $D^{0,+}$, $D_{s}^{+}$ and $D^{*+}$ mesons and spectrum ratios in central rapidity region in low $p_{T}$ range $p_{T}\lesssim7$ GeV/$c$ in minimum-bias events are well described by the quark combination mechanism. We emphasized two important foundations in calculations: (1) the $p_{T}$ spectra of light-flavor quarks in the low $p_{T}$ range are extracted from the data of light-flavor hadrons using quark combination mechanism also in equal-velocity combination approximation; (2) the used $p_{T}$ spectrum of charm quarks is consistent with perturbative-QCD calculations. 

The $\Lambda_{c}^{+}/D^{0}$ ratio in quark combination mechanism exhibits a typical increase-peak-decrease behavior as the function of $p_{T}$. The shape of the ratio for $p_{T}\gtrsim3$ GeV/$c$ is in agreement with the data of ALICE collaboration in the central rapidity region and those of LHCb collaboration in the forward rapidity region. The enhanced production of single-charm baryons is parameterized by $R_{B/M}^{\left(c\right)}$ in the model, which is found to be quite high at central rapidities by fitting ALICE data, i.e., $R_{B/M}^{\left(c\right)}\approx0.43$. It is significantly larger than that from the empirical extrapolation from light flavor hadrons ($R_{B/M}^{\left(c\right)}\approx0.26$). The $R_{B/M}^{\left(c\right)}$ extracted from the preliminary data of $\Lambda_{c}^{+}/D^{0}$ ratio in forward rapidity region reported by LHCb collaboration is about 0.29, which is close to the extrapolation from light flavor hadrons.  We made discussions on the possible reasons of the enhancement of $R_{B/M}^{\left(c\right)}$ existing in literature. 

We made predictions on the $p_{T}$ spectra of $\Xi_{c}^{0}$ and $\Omega_{c}^{0}$ baryons and their ratios to $D_{s}^{+}$ in central rapidity region to make further test of such significant enhancement of charm baryons.  In particular, the $\Xi_{c}^{0}/D_{s}^{+}$ ratio shows the similar enhancement with $\Lambda_{c}^{+}/D^{0}$. In addition, we predicted the production of $D$ mesons, $\Lambda_{c}^{+}$, $\Xi_{c}^{0}$ and $\Omega_{c}^{0}$ baryons in high multiplicity classes in $p$-Pb collisions at $\sqrt{s_{NN}}=5.02$ TeV for the future test. 

\begin{acknowledgments}
The authors thank Qun Wang for helpful discussions. This work is supported by the National Natural Science Foundation of China under Grant Nos.  11575100, 11505104.
\end{acknowledgments}

\bibliographystyle{apsrev4-1}
\bibliography{ref}

\appendix
\section{ Estimation of $R_{V/P}$, $R_{S3/S1}$ and $R_{S1/T}$ by effective statistical weights \label{appendixA} }

In the combination of a given $c\bar{l}$ or a $cll$, the formed meson or baryon can have different spin states.  We use an effective statistical model to roughly estimate the relative production weight of different spin states. We recall the fact that the data of multiplicities of light-flavor hadrons and open charm hadrons in elementary $e^{+}e^{-}$, $pp$ and $p\bar{p}$ reactions were shown to be well described by the statistical model \cite{Becattini:1997rv,Andronic:2009sv}, which indicates the statistical description of hadron multiplicity captures the key characteristics of the non-perturbative hadronization \cite{Stock:2007gh, Castorina:2007eb,Stock:1999hm,Heinz:2000ba,Hatta:2008qx}. 

    In the Boltzmann limit of Bose and Fermi statistics, the hadron multiplicity is the phase space integral of Boltzmann distribution, 
\begin{align}
    N &= (2J+1) \frac{V}{(2\pi)^3} \int d^{3}p\, e^{-\sqrt{p^2+m^2}/T } \label{a1} \\
        &= (2J+1) \frac{V T}{2\pi^2} m^{2}  K_{2}\left(\frac{m}{T}\right), \label{a2}
\end{align}
where $m$ is mass and $J$ is spin. $V$ is the proper volume of the system and $T$ is temperature. $K_2$ is the modified Bessel function of order 2. 

Using Eq.~(\ref{a2}), we get the ratio of vector meson $D^*$ ($m_{D^*}=2.01$ GeV) to pseudo-scalar meson $D$ ($m_D=1.87$ GeV)
\begin{equation}
    \frac{D^{*}}{D} = 3\frac{m_{D^*}^{2} K_{2}(m_{D^*}/T)}{m_D^2 K_{2}(m_D/T)}\approx 1.45,
\end{equation}
and that of $D_{s}^{*}$ ($m_{D_{s}^{*}}=2.11$ GeV) to $D_s$ ($m_{D_s}=1.97$ GeV), 
\begin{equation}
    \frac{D_s^{*}}{D_s}  \approx 1.45
\end{equation}
at temperature $T=170$ MeV \cite{Andronic:2009sv}. We take the above statistical weights as a guideline and take $R_{V/P}=1.5$ in our calculation. 

For baryons, we have 
\begin{equation}
    \frac{\Sigma_{c}}{\Lambda_{c}^{+}} \approx 0.41,  \hspace{2cm} \frac{\Xi'_{c}}{\Xi_{c}} \approx 0.57,
\end{equation}
with $m_{\Sigma_c}=2.46$ GeV, $m_{\Lambda_{c}^{+}}=2.29$ GeV, $m_{\Xi'_c}=2.58$ GeV, and $m_{\Xi_c}=2.47$ GeV.  We take the average of above two statistical weights as a guideline and take $R_{S1/T}=0.5$ in our calculation.

For sextet baryons of spins $1/2$ and $3/2$, we have 
\begin{align}
    \frac{\Sigma_{c}^{*}}{\Sigma_{c}} \approx 1.41, \hspace{1cm} \frac{\Xi_{c}^{*}}{\Xi'_{c}} \approx 1.38, \hspace{1.cm} \frac{\Omega_{c}^{*}}{\Omega_{c} } \approx 1.37, 
\end{align}
with $m_{\Sigma_{c}^{*}}=2.52$ GeV,  $m_{\Xi_{c}^{*}}=2.65$ GeV, $m_{\Omega_{c}^{*}}=2.77$ GeV, $m_{\Omega{c}}=2.70$ GeV. We take the average of above  statistical weights as a guideline and take $R_{S3/S1}= 1.4$ in our calculation.

\section{$R_{B/M}^{(c)}$ extrapolated from light-flavor hadrons \label{appendixB}}

For simplicity, we use $l$ to denote light-flavor quarks and $c$ charm quarks, and $N_{l}$ and $N_{c}$ their numbers at hadronization.  $N_{q}=N_{l}+N_{c}$ is the total number of quarks. The numbers of the formed light-flavor baryons ($B_{lll}$), single-charm baryons ($B_{cll}$), multiple-charm baryons ($B_{ccl}$ and $B_{ccc}$) are $N_{B_{lll}}$, $N_{B_{cll}}$, $N_{B_{ccl}}$ and $N_{B_{ccc}}$.
The total number of baryons is 
\begin{equation}
    N_{B}=N_{B_{lll}}+N_{B_{cll}}+N_{B_{ccl}}+N_{B_{ccc}}.
    \label{nb_nor}
\end{equation}
If we assume the flavor-blind baryon formation probability, we have
\begin{align}
    N_{B_{f_{1}f_{2}f_{3}}} & =N_{f_{1}f_{2}f_{3}} P_{f_{1}f_{2}f_{3}\to B}, \label{Nb_f1f2f3}\\
P_{f_{1}f_{2}f_{3}\to B} & =N_{iter,f_{1}f_{2}f_{3}} \frac{N_{B}}{N_{qqq}},
\end{align}
where indexes $f_{1},f_{2},f_{3}=l,c$. The coefficient $N_{iter,f_{1}f_{2}f_{3}}$ is (1, 3, 3, 1) for cases ($lll$, $cll$, $ccl$, $ccc$), respectively, so that $N_{lll}+3N_{cll}+3N_{ccl}+N_{ccc}=N_{qqq}$ and the normalization Eq. (\ref{nb_nor}) is satisfied. Then we get 
\begin{equation}
    P_{cll\to B_{cll}}=3 P_{lll\to B_{lll}}. \label{Pcll}
\end{equation}

Similarly, the numbers of the formed light-flavor mesons ($M_{l\bar{l}}$), single-charm mesons $M_{c\bar{l}}$ and $M_{\bar{c}l}$, and charmonium $M_{c\bar{c}}$ are $N_{M_{l\bar{l}}}$, $N_{M_{c\bar{l}}}$, $N_{M_{\bar{c}l}}$, and $N_{M_{c\bar{c}}}$. The total meson number is $N_{M}=N_{M_{l\bar{l}}}+N_{M_{c\bar{l}}}+N_{M_{\bar{c}l}}+N_{M_{c\bar{c}}}$.
In flavor-blind approximation, we have 
\begin{align}
    N_{M_{f_{1}\bar{f}_{2}}} & =N_{f_{1}\bar{f}_{2}} P_{f_{1}\bar{f}_{2}\to M}, \label{Nm_f1f2} \\
P_{f_{1}\bar{f}_{2}\to M} & =\frac{N_{M}}{N_{q\bar{q}}},
\end{align}
where indexes $f_{1},f_{2}=l,c$. Then we get 
\begin{equation}
    P_{c\bar{l}\to M_{c\bar{l}}}=P_{l\bar{l}\to M_{l\bar{l}}}. \label{pcl}
\end{equation}
Using Eqs.~(\ref{Nb_f1f2f3}), (\ref{Pcll}), (\ref{Nm_f1f2}) and (\ref{pcl}), the ratio of single-charm baryons to single-charm mesons is 
\begin{align}
R_{B/M}^{\left(c\right)} & =\frac{N_{B_{cll}}}{N_{M_{c\bar{l}}}}=3\frac{N_{cll}}{N_{c\bar{l}}}\frac{P_{lll\to B_{lll}}}{P_{l\bar{l}\to M_{l\bar{l}}}}\\
    &=3\frac{N_{cll}}{N_{c\bar{l}}}\frac{N_{l\bar{l}}}{N_{lll}}\frac{N_{B_{lll}}}{N_{M_{l\bar{l}}}}\\
 & =3\frac{N_{l}}{N_{l}-2}\frac{N_{B_{lll}}}{N_{M_{l\bar{l}}}} \\
    &\approx3\frac{N_{B_{lll}}}{N_{M_{l\bar{l}}}}=3R_{B/M}^{\left(l\right)}
\end{align}

\end{document}